\documentclass[12pt,preprint]{aastex}
\usepackage{graphicx,picins,wrapfig}           
\usepackage[ngerman,english]{babel}
\usepackage[T1]{fontenc}
\usepackage[ansinew]{inputenc}
\usepackage{floatflt}           
\usepackage{rotating}   
\usepackage{placeins}
\usepackage{ae}
\usepackage{amsmath,amssymb,amstext,amscd}
\usepackage{natbib}
\usepackage[font=scriptsize]{subfig}            
\usepackage{times}
\usepackage{lettrine}
\usepackage{fancyhdr}           
\newcommand{\myemail}{klement,rix@mpia.de}
\newcommand{\fuchsemail}{fuchs@ari.uni-heidelberg.de}
\newcommand{\chrisemail}{cflynn@utu.fi}
\newcommand{\beersemail}{beers,lee@pa.msu.edu}
\newcommand{\paolaemail}{paola.refiorentin@fmf.uni-lj.si}
\newcommand{\observersemail}{snedden, dmbiz, hjbrew, viktorm, elenam, doravetz, kpan, asimmons@apo.nmsu.edu}
\newcommand{\carlosemail}{callende@astro.as.utexas.edu}
\defcitealias{kle08}{Paper~I}

\begin{document}

\title{Halo Streams in the 7$^{th}$ SDSS Data Release}

\author{R. Klement\altaffilmark{1}, H.-W. Rix\altaffilmark{1}, C.
Flynn\altaffilmark{2}, B. Fuchs\altaffilmark{3}, T. C.
Beers\altaffilmark{4}, C. Allende Prieto\altaffilmark{7,8}, D. Bizyaev\altaffilmark{6}, H. Brewington\altaffilmark{6}, Y. S. Lee\altaffilmark{4}, E. Malanushenko\altaffilmark{6}, V. Malanushenko\altaffilmark{6}, D. Oravetz\altaffilmark{6}, K. Pan\altaffilmark{6}, P. Re Fiorentin\altaffilmark{1,5}, A. Simmons\altaffilmark{6}, S. Snedden\altaffilmark{6}}

\altaffiltext{1}{Max-Planck-Institut f\"ur Astronomie, K\"onigstuhl 17, D-69117 Heidelberg; \myemail}
\altaffiltext{2}{Tuorla Observatory, V\:ais\:al\:antie 20, FI-21500 PIIKKI\:O; \chrisemail}
\altaffiltext{3}{Astronomisches Rechen-Institut am Zentrum f\"ur Astronomie Heidelberg, M\"onchhofstraße 12-14, D-69120 Heidelberg; \fuchsemail}
\altaffiltext{4}{Department of Physics \& Astronomy, CSCE: Center for the Study
of Cosmic Evolution, and JINA: Joint Institute for Nuclear Astrophysics, Michigan State University, E. Lansing, MI 48824, USA; \beersemail}
\altaffiltext{5}{Department of Physics, University of Ljubljana, Jadranska 19, 1000 Ljubljana Slovenia; \paolaemail}
\altaffiltext{6}{Apache Point Observatory, Sunspot, NM 88349, USA; \observersemail}
\altaffiltext{7}{McDonald Observatory and Department of Astronomy, The University of Texas, 1 University Station, C1400, Austin, TX 78712-0259, USA; \carlosemail}
\altaffiltext{8}{Mullard Space Science Laboratory, University College London, Holmbury St. Mary, Surrey RH5 6NT, United Kingdom }

\begin{abstract}

We have detected stellar halo streams in the solar neighborhood using data
from the 7$^{th}$ public data release of the Sloan Digital Sky Survey (SDSS), which
includes the directed stellar program SEGUE: Sloan Extension For Galactic
Understanding and Exploration. In order to derive distances to each star, we used the
metallicity-dependent photometric parallax relation from \citet{ive08}.
We examine and quantify the accuracy of this relation by
applying it to a set of globular and open clusters observed by SDSS/SEGUE, and
comparing the resulting sequence to the fiducial cluster sequences obtained by
\citet{an08}. Our final sample consists of 22,321 nearby ($d\leq2$ kpc),
metal-poor ([Fe/H] $\leq-0.5$) main-sequence stars with 6D estimates of position
and space velocity, $(\vec{r},\vec{v})$. We characterize the orbits of these
stars through suitable kinematic proxies for their ``effective'' integrals of
motion, angular momentum, eccentricity, and orbital polar angle and compare the
observed distribution to expectations from a smooth distribution in four
[Fe/H] bins. The metallicities provide an additional dimension in parameter
space that is well suited to distinguish tidal streams from those of dynamical
origin. On this basis we identify at least five significant ``phase-space
overdensities'' of stars on very similar orbits in the solar neighborhood to
which we can assign unambiguously peaked [Fe/H] distributions. Three of them
have been identified previously, including the halo stream discovered by Helmi et
al. (1999) at a significance level of $\sigma=12.0$. In addition, we find at least two new genuine halo streams, judged
by their kinematics and [Fe/H], at $\sigma=2.9$ and 4.8, respectively. For one stream the stars even show coherence in configuration space, matching a 
spatial overdensity of stars found by \citet{jur08} at $(R,z)\approx(9.5,0.8)$ kpc.
Our results demonstrate the practical power of our
search method to detect substructure in the phase-space distribution of nearby
stars {\em without} making a-priori assumptions about the detailed
form of the gravitational potential.

\end{abstract}

\keywords{Galaxy: solar neighborhood --- Galaxy: kinematics and dynamics}

\section{Introduction}

The phase-space distribution of stars in the solar neighborhood encodes enormous
amounts of information on the present dynamical state and the formation history
of the Milky Way. A key role in this context is played by the existence of
substructure in the phase-space distribution of stars, caused by stellar streams
or moving groups, that is, groups of stars moving on similar orbits in the Milky
Way's gravitational potential. Such moving groups have been known to exist in
the velocity distribution of nearby stars for some time \citep[and references
therein]{pro1869,lin25,egg96}. Stellar streams, together with their chemical and
dynamical properties, can be used to constrain various scenarios of the
hierarchical build-up of the Milky Way \citep[e.g.,][]{helm99} as well as its
gravitational potential \citep[e.g.,][]{ant08}. 

Moving groups in the solar neighborhood emerge for several reasons.
The simplest case is an agglomeration of stars that were born in
the same molecular cloud and only recently dissolved. In this case the stars
keep on moving in the direction of the once-bound cluster, until phase-mixing
washes out their common orbital signature. While this scenario seems to be valid for
some stellar streams \citep[e.g., HR1614,][]{desilv07}, the chemical and
chronological properties of the largest moving groups are incompatible with it.
For example, \citet{cher01} reported an age range of 0.5 Gyr to more than 2-3
Gyr for the Hyades supercluster, along with a rather large velocity dispersion, which
they identified with the presence of several subgroups. Such subgroups have also
been found by \citet{dehn98}, who used subsamples of young and old stars based
on their spectral types. He discovered an asymmetric drift relation for the
moving groups, in the sense that those only present in the red subsamples (old stars)
have larger radial velocity components, and lag with respect to the Local
Standard of Rest, than those also containing blue (younger) stars. In other
words, old moving groups move on more eccentric orbits. To explain this
observation, \citet{dehn98} proposed that these streams consist of stars that have been
trapped onto nearly resonant orbits that oscillate about their parent resonant
orbits, while the latter slowly change their eccentricity along with the
non-axisymmetric potential. This interpretation is based on 
suggestions already made by \citet{may72} and \citet{kal91}. The latter tried to
explain the bimodal velocity distribution of the Sirius (moving radially
inwards) and Hyades (moving radially outwards) streams by putting the Sun at the
position of the outer Lindblad resonance (OLR) of the Galactic bar. However,
\citet{fam04} later pointed out that these streams are better explained as stars
on horseshoe orbits that cross-corotate
in the rest frame of spiral density waves \citep[for more details, see][]{sel02}. They further argued that the
clusters of coeval stars that have traditionally been connected to these streams
would have been picked up by the spiral waves along with field stars of
different ages, and therefore are just moving in these kinematic groups by chance.
\citet{desim04} also found that spiral waves can produce kinematic structures
similar to those observed in the solar neighborhood, although they attributed
this more to disk heating rather than radial migration. 

\citet{deh00} and \cite{fux01} later used the position of the Hercules stream,
which is lagging the Local Standard of Rest by $\sim50$ km s$^{-1}$, to
constrain the inclination angle and position of the OLR of the Galactic bar.
Similarly, \citet{qui05} found that placing the Sun near the 4:1 inner Lindblad
resonance with a two-armed spiral density wave could account for the
velocity-space positions of the Hyades and Coma Berenices moving groups. Very
recently, simulations of the birth and evolution of disk stars in a Milky Way
potential including axisymmetric components for the disk, the bulge and halo,
spiral arms, and a bar, were able to reproduce the shape of the Hercules, Coma
Berenices, Hyades and Sirius moving groups in velocity space \citep{ant08}. These examples show that the velocity distribution, as well as the age and
chemical composition of dynamical streams in the solar neighborhood, can be used
as tracers of the Galactic potential. 

A third scenario for the formation of a stellar stream is a tidally disrupting
cluster or satellite galaxy that deposits its debris on similar orbits (that is,
a ``tidal'' or ``halo stream''). \citet{helm99a} performed simulations of
disrupting satellites crossing the solar neighborhood, and showed that the debris
loses its spatial coherence completely over a Hubble time. In contrast,
the stream stars clump together in velocity space, resembling classical moving groups.
The reason is that stars in a single stream obey the collisonless Boltzmann
equation, a special case of Liouville's theorem, which states that the phase-space
density of a stellar sub-population is conserved at any given phase-space point.
Initially, the stream stars are located in a small phase-space volume.
However, as the spatial components of the stars disperse with time, they 
become more focused in their velocity components. Although in practice the
velocity dispersion of a tidal stream tends to increase with time, due to the
effect of phase-mixing \citep{helm99a}, such tidal streams will still form
coherent features in velocity space, because to be in the solar neighborhood at
the same time the azimuthal velocities of the stars must be similar. This has
been confirmed with a number of different data sets, and stellar halo streams
have been identified in the kinematic distribution of solar neighborhood stars
\citep[hereafter Paper~I]{helm99,chi00,nav04,ari06,helm06,det07,kle08}. Tidal
streams also conserve the so-called integrals of motion of their progenitor,
energy and angular momentum, allowing their recovery even if the halo has
undergone complete mixing \citep{helm00}. Trying to confine Milky Way streams
into a small range of energies could yield a best-fit to the gravitational
potential.

The primary goal of this paper is to search for substructure in the solar
neighborhood that can be attributed to tidal halo streams. \citet{helm99}
predicted that $\sim500$ kinematically cold streams might exist in the solar
neighborhood, yet only a few have been detected so far. Currently, however, the
Sloan Digital Sky Survey \citep[SDSS,][]{york00} and its extensions, which have
thus far obtained spectra for over 400,000 stars, represents the most extensive database 
collected to date to increase the number of detected halo streams. As has been shown
by simulations from various authors \citep[e.g.,][]{helm00,pen06,choi07}, a
straightforward approach to find stellar streams would be in the space of the
integrals of motion $(E,L,L_z)$, which are immune to phase-mixing. In practice,
however, the integrals of motion are not uniquely defined, because the potential
is not exactly known. Furthermore, the error bars on 6D measurements are drastically
``anisotropic'' across the different components. We therefore want to explore search
strategies in a modified integrals-of-motion-space that match the data and error
bars. 

This paper is organized as follows. Section~\ref{sec2} describes our dataset,
our methods for deriving distances, comparisons of our adopted photometric
parallax relation with cluster fiducial sequences, and estimates of
space velocities for the stars.  Section~\ref{sec3} describes our strategy for
searching for streams in the solar neighborhood.  We apply our methods in
\S~\ref{sec4}, pointing out where several of our streams overlap with previously
detected examples.  We consider the potential effects of systematic distance
errors in \S~\ref{sec:systdisterr}. Section~\ref{sec6} discusses our methods for the
determination of the statistical significance associated with stream detection.
In \S~\ref{sec7} we present four new likely streams, and confirm the
detection of three others identified in previous work (adding additional members
to these structures).  Our conclusions and a brief discussion are presented in
\S~\ref{sec8}. 

\section{The Data}\label{sec2}

SDSS-I was an imaging and spectroscopic survey that began routine operations
in April 2000, and continued through June 2005. The SDSS, and its extensions,
uses a dedicated 2.5m telescope \citep{gunn06} located at the Apache Point
Observatory in New Mexico. The telescope is equipped with an imaging camera and
a pair of spectrographs, each of which are capable of simultaneously collecting
320 medium-resolution ($R = 2000$) spectra over the seven square degree field of
view, so that on the order of 600 individual target spectra and roughly 40
calibration-star and sky spectra are obtained on a given spectroscopic
``plug-plate'' \citep{york00}. The imaging camera \citep{gunn98} contains an
imaging array of 30 4-megapixel CCDs and astrometric arrays that measure fluxes
for calibration with standard astrometric catalog stars. The flux densities of
observed objects are measured almost simultaneously in five bands [$u,g,r,i,z$],
with effective wavelengths of [3540~{\AA}, 4760~{\AA}, 6280~{\AA}, 7690~{\AA},
9250~{\AA}], respectively \citep{fuk96,gunn98,hogg01}. The camera sweeps the sky
in great circles (in drift scan mode) and a point on the sky passes the filters
in the order of $r,i,u,z,g$. The brightness limit where the imaging camera
saturates is at $g\sim14$ mag. The completeness at this magnitude is
$\sim99.3$\% for point sources \citep{ive01}; it drops to 95\% at magnitudes of
[22.1, 22.4, 22.1, 21.2, 20.3]\footnote{These values have been derived by
comparing multiple scans of the same area obtained during the commissioning year
with typical seeing of $1.5"\pm0.1"$.}. The SDSS photometry is accurate to 0.02
mag rms at the bright end\footnote{This value is determined using repeated
observations of 3,000,000 point sources over time spans ranging from 3 hours to
3 years.}, with well controlled tails of the error distribution \citep{ive03}.
Astrometric positions are accurate to about 0.1" per coordinate for sources
brighter than $r\sim20.5$ mag \citep{pier03}. Morphological information from the
images allows point source-galaxy separation to $r\sim21.5$ mag \citep{lup02}. 

One of three sub-surveys carried out during the first extension of SDSS, 
known as SDSS-II, the Sloan Extension for Galactic Understanding and
Exploration (SEGUE), ran from July 2005 to June 2008. SEGUE obtained some
250,000 medium-resolution spectra of stars in the Galaxy, selected in order to
explore the nature of stellar populations from 0.5 kpc to 100 kpc \citep{yan09}. 
These data, along with the substantial numbers of suitable stars observed
during the course of SDSS-I, permit the derivation of the full six-dimensional
phase-space distribution of the various components of the Milky Way. Stellar
physical parameters (T$_{\rm eff}$, log g, [Fe/H]), based on SDSS photometry and
spectroscopy, are derived by application of the SEGUE Stellar Parameter Pipeline
(SSPP) described by Lee et al. (2008a,b) and \citet{alle08}.

\begin{figure}[!ht]
\centering
 \includegraphics[width=\linewidth]{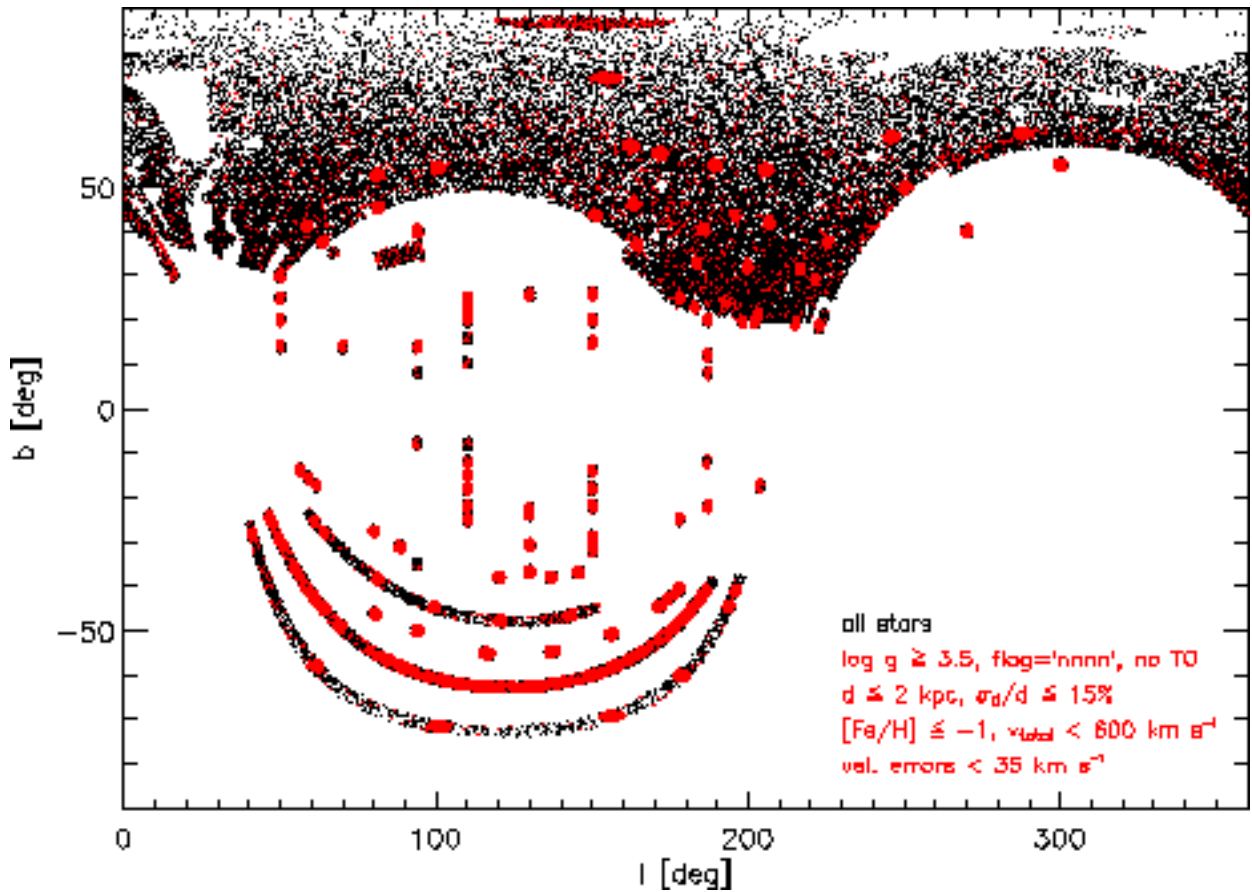}
   \caption[Sky coverage of our sample of SDSS/SEGUE stars]{Sky coverage of our
sample of SDSS/SEGUE stars. Each star is plotted individually. The red data points
represent the positions of metal-poor stars that meet the selection criteria of
our final sample (see \S~\ref{sec:FinalSample}). 
Note the sparse sampling obtained during SDSS-I vs. 
the focused sampling of the SDSS-II/SEGUE pointings.\label{fig:SDSS-l-b}}
\end{figure}

We start the sample selection for the present study from all stars targeted for
spectroscopy by SDSS/SEGUE with S/N$>10$, accepted photometry in all five bands, and
estimates for the radial velocity, [Fe/H], and proper motions. Note that the
requirement for determinations of [Fe/H] is essentially one on effective
temperature (or color), as the present SSPP provides confident estimates of
metallicity over the range 4500~K $\leq$ T$_{\rm eff} \leq$ 7500~K. These stars
have been taken from the seventh data release \citep{aba09}; their proper
motions have been corrected for a systematic error that occured in the data
reduction procedure (Munn 2008, internal SDSS memorandum). There are a number of
repeated observations, either for quality assurance, or from re-use of
photometric calibration stars by several plug-plates. These repeats are
independent observations and are listed separately in the SDSS Catalog Archive
Server, with different identification numbers. We only keep one object per
position on the sky, to which we assign a radial velocity and stellar parameters
averaged over all repeats. Figure~\ref{fig:SDSS-l-b} shows the sky coverage of
the resultant sample of 154,888 stars. These data cover a large, almost
contiguous area in the Northern Galactic Cap, plus three stripes in the South
Galactic Cap.

\subsection{Distance Estimates}

The majority of stars spectroscopically targeted by SDSS-I are main-sequence stars
(or metal-poor main-sequence turnoff stars used as calibration objects)
\citep[$\sim99\%$,][]{fin00}. Although the targets spectroscopically
selected by SEGUE explicitly include giants, their fraction (based on
spectoscopic surface gravity estimates) remains low. Only 8.8\% of these stars
have $\log \text{g}<3.5$, the surface gravity where we divide between dwarfs and
giants. This is a bit more stringent than the separation made, e.g., by
\citet{ive08} at $\log \text{g}=3$, but we wish to ensure that only
late-type dwarfs and subdwarfs are selected.\footnote{A subdwarf is defined as a
star with luminosity 1.5 to 2 magnitudes lower than that of a Solar-metallicity main-sequence star
of the same spectral type.} Presuming a sample dominated by main-sequence stars,
we can apply a photometric parallax relation to derive distances. Because we
want to concentrate on a wide range of metal-poor stars, the effect of
metallicity on the absolute magnitude at a given color becomes important. We
have spectroscopic metallicities available for each star, so we are motivated to
adopt a photometric parallax relation that explicitly accounts for metallicity
over a wide range of colors.

Such a relation has been derived by \citet{ive08}. The shape of their
color-magnitude relation, $M_r^0(g-i)$, is constrained by simultaneously fitting
SDSS photometry data for five globular clusters, normalized to the same
arbitrary magnitude scale by requiring the same median 
magnitude ($r=0$) for stars in the color range $0.5 < g-i < 0.7$. By assuming
that this shape depends only on color, not metallicity, and its normalization
depends only on metallicity, not color, the absolute magnitude offset of each
cluster from the mean relation can be expressed as a function of metallicity.
The absolute magnitude of a star is then calculated as:

\begin{equation}\label{eq:SEGUE01}
	M_r(g-i,\text{[Fe/H]})=M_r^0(g-i)+\Delta M_r(\text{[Fe/H]}).
\end{equation}

With distances adopted from \citet{har96} and six additional open and globular
cluster data from \citet{van03}, they derive the following absolute magnitude
correction:

\begin{equation}\label{eq:SEGUE02}
	\Delta M_r(\text{[Fe/H]})=4.50-1.11\text{[Fe/H]}-0.18\text{[Fe/H]}^2
\end{equation}

The correction~\eqref{eq:SEGUE02} suggests an offset from the mean relation of 4.5
for solar metallicity stars, due to the scaling to $r=0$ described above. 
\citet{ive08} further expand the mean photometric parallax relation to the color
range $0.2 < g-i < 4.0$ by using constraints from trigonometric parallaxes given
in Bochanski et al. (2008, in prep.), additional cluster data observed in the
SDSS from \citet{clem08}, and an age correction for turnoff stars. The result is
a fifth-order polynomial:

\begin{equation}\label{eq:SEGUE03}\begin{split}
	M_r^0(g-i)=&-5.06+14.32(g-i)-12.97(g-i)^2\\
	&+6.127(g-i)^3-1.267(g-i)^4+0.0967(g-i)^5,
	\end{split}
\end{equation}

\noindent which, together with Equations~\eqref{eq:SEGUE01} and
\eqref{eq:SEGUE02}, is our adopted photometric parallax relation.

We test the validity of this relation for our sample using different approaches,
as described below.

\subsubsection{Comparison of the Photometric Parallax Relation with Cluster Fiducial Sequences}
\label{sec:ComparisonWithFiducials}

An et al. (2008, hereafter An08) have used crowded-field photometry
techniques to analyse SDSS/SEGUE imaging data for 17
globular and 3 open clusters, and determined fiducial sequences from their
color-magnitude diagrams (CMDs). These sequences give the $r$-band magnitude as a
function of either $u-g$, $g-r$, $g-i$, or $g-z$ color. This is the first time
that cluster fiducial sequences have been evaluated in the native SDSS $ugriz$
system, allowing for tests of the photometric parallax relation from
\citet{ive08} without the need to rely on color transformations from other
systems. 

\begin{figure}[!htb]
\centering
 \includegraphics[width=\linewidth]{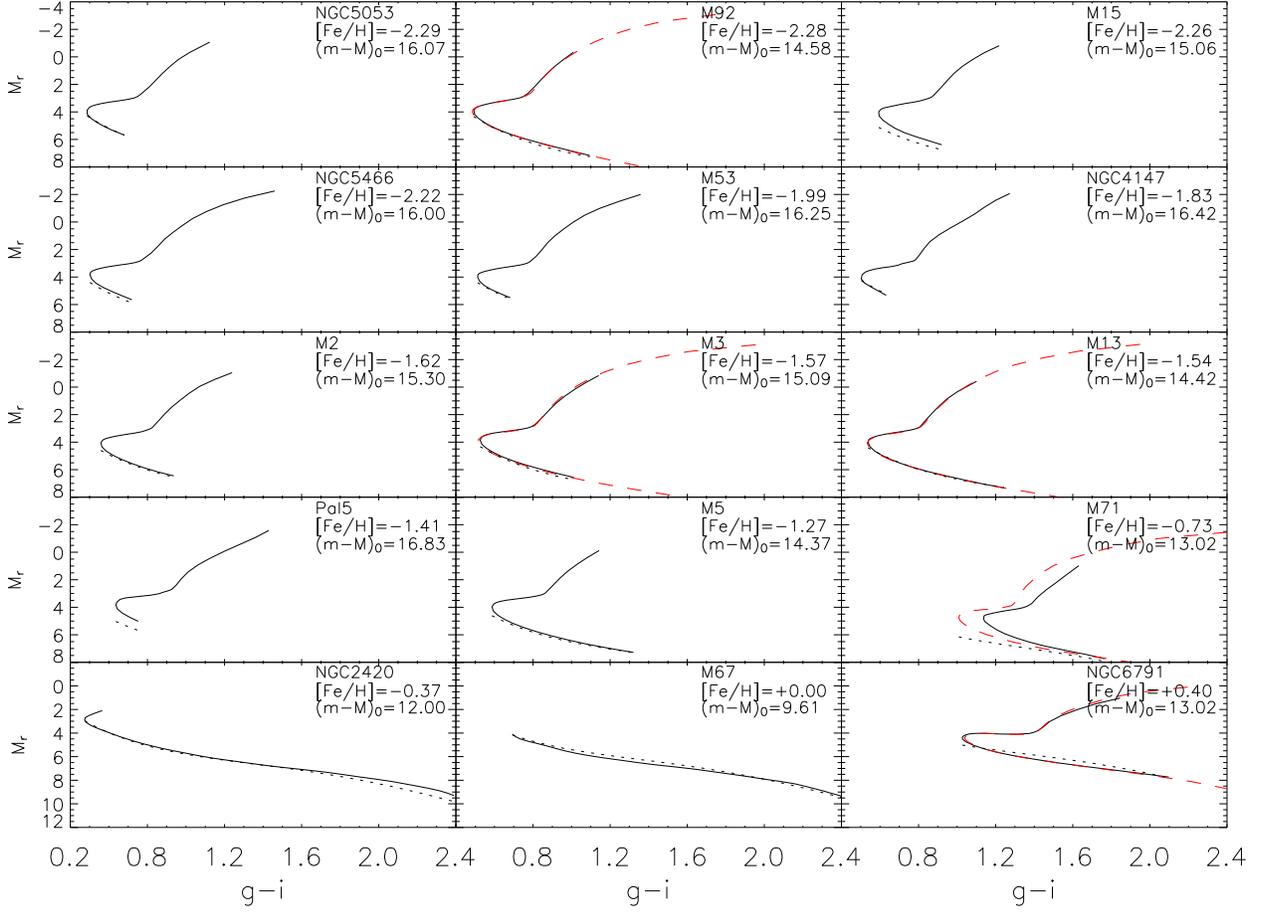}
   \caption[Comparison of the \citet{ive08} photometric parallax relation to
   fiducial sequences]{Comparison of cluster fiducial sequences taken from An08
   (\textit{solid black lines}) and \citet{clem08} (\textit{dashed red line}),
   respectively, and the [Fe/H]-dependent photometric parallax relation from
   \citet{ive08} (\textit{dotted line}). The panels show the absolute magnitude,
   $M_r$, as a function of the $g-i$ color. The adopted distance moduli and
   metallicities are taken from various sources in the literature (see An08,
   \S~2, for all references).}
\label{fig:f2}%
\end{figure}

We use 15 of the cluster fiducials and compare them to the sequences derived
from Equations~\eqref{eq:SEGUE01}-\eqref{eq:SEGUE03}. In addition, we consider the
fiducial sequences from \citet{clem08} for five clusters, after transforming them
onto the $ugriz$ system using the transformations given by \citet{tuck06}. These
sequences have been shown to match the An08 fiducials within the errors of the
photometric zero points. Because they were obtained from observations with
various integration times, the Clem et al. sequences extend over a broader
magnitude range than the An08 sequences. We calculated the absolute magnitude,
$M_r$, for each cluster by adopting the distance modulii and metallicities as given
in An08. The results are shown in Figure~\ref{fig:f2}, sorted by decreasing
cluster metallicity. 

For most of the clusters, the photometric parallax relation of \citet{ive08}
fits the fiducial main sequences remarkably well. In the case of nearly all
the metal-poor clusters with [Fe/H]$<-1.0$, the difference between the absolute
magnitude predicted by the photometric parallax relation, M$_{r,\text{phot}}$,
and the absolute magnitude given by the cluster fiducial sequences, M$_{r,
\text{cluster}}$, stays below $\sim$0.2 mag for $g-i\gtrsim0.4$. There are three
exceptions to this trend.  One is M15, where the turnoff is slightly redwards
of $g-i=0.4$ and the absolute magnitude offset drops roughly from M$_{r,
\text{phot}}$-M$_{r,\text{cluster}}$$=$0.48 mag at $g-i=0.51$ to 0.36 mag at
$g-i=0.77$. For Palomar 5, the discrepancy is more than 0.7 mag across the main
sequence. However, this cluster is known to be in the process of tidal
disruption \citep{oden01,grill06} and is sparsely populated in the observations;
contamination by foreground and background stars is possible. Also, the color
range spanned by the main sequence is very small compared with the giant
branch, and it may not extend far enough from the turnoff for the photometric
parallax relation to be valid. The An08 fiducial sequence for M71 has to be taken
with caution, because according to these authors the zero points for the M71
photometry were very uncertain, and there was a strong contamination by
likely background stars. The uncertain fiducial sequence could thus account for
the offsets.

\begin{figure}[!htb]
\centering
 \includegraphics[width=\linewidth]{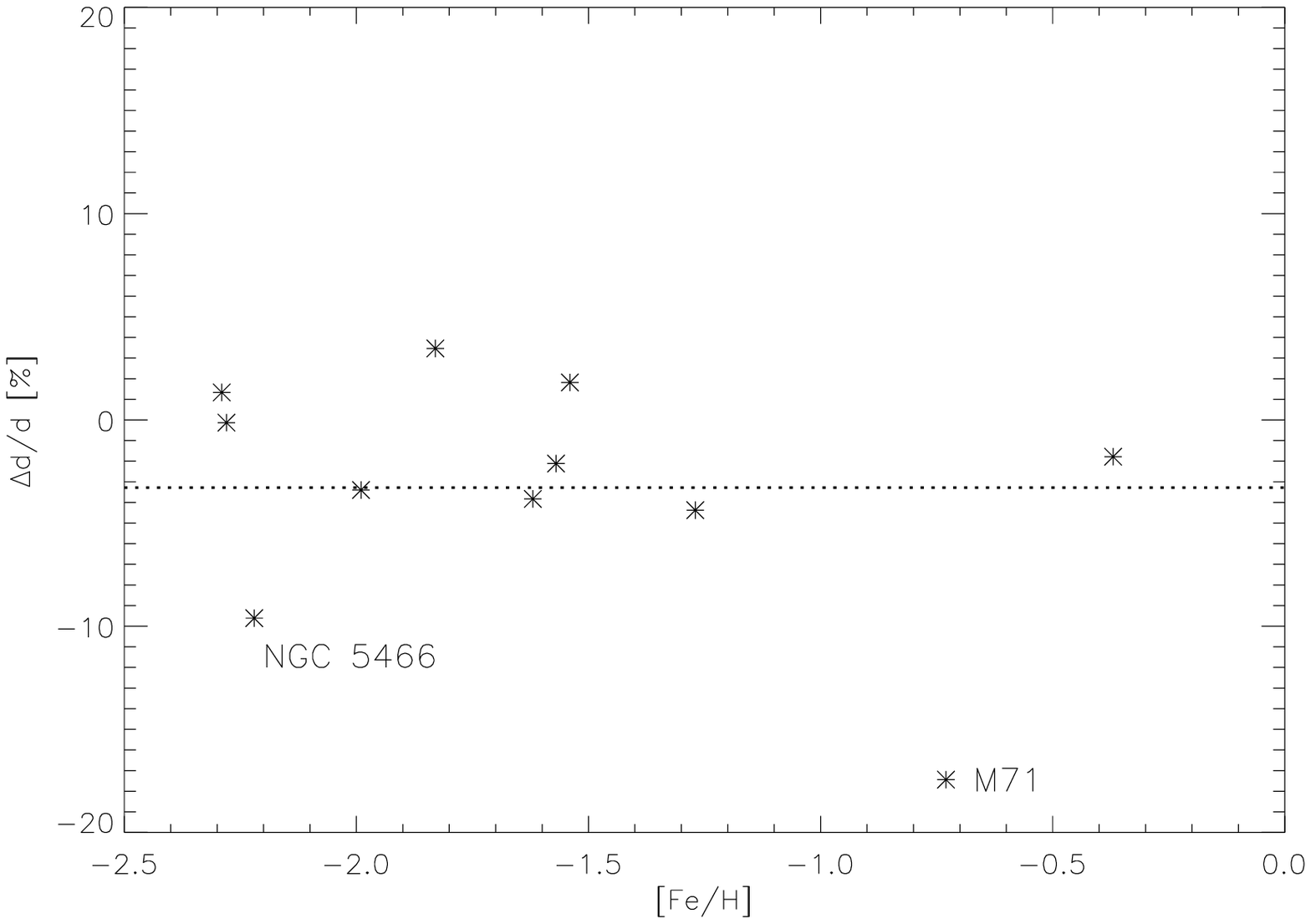}
   \caption[Histogram of systematic distance errors for eleven clusters with
   $-2.0<\text{[Fe/H]}<-0.3$]{Relative distance errors, $\Delta
   d/d=(d_\text{phot}-d_\text{cluster})/d_\text{cluster})$, vs. metallicity,
   [Fe/H], derived from eleven of the 13 clusters that lie in the metallicity
   range $-2.0<\text{[Fe/H]}<-0.3$. The errors of each cluster have been
   averaged over the color range $(g-i)_\text{TO}+0.05<g-i<1.5$. The two
   clusters M15 and Pal5 have not been considered, while M71 has been kept,
   because it has a sequence measured by \citet{clem08} with deep photometry.
   The dotted line indicates the distance offset averaged over all eleven
   clusters (including M71), which we adopt as the systematic distance error for
   the \citet{ive08} photometric parallax relation.}
\label{fig:f3}%
\end{figure}

We find it more descriptive to express the systematic differences between the
photometric parallax relation and the cluster fiducial sequences through a
distance offset rather than a magnitude offset. To derive an estimate of the
systematic distance error of the photometric parallax relation, we average over
the differences between the distance that is predicted by the relation,
$d_\text{phot}$, and the distance given by the distance modulus of each cluster,
$d_\text{cluster}$. To this end, we concentrate on the color range $(g-i)
_\text{TO}+0.05<g-i<1.6$, where $(g-i)_\text{TO}$ denotes the turnoff color,
that is, the color at which a cluster's main sequence runs
vertically.\footnote{This is the bluest color for which a value of the fiducial
sequence exists.} The color $g-i=1.6$ corresponds approximately to the reddest color in
our sample. If available, we prefer the cluster sequences of \citet{clem08} for
calculating $\Delta d/d$. After elimination of the two outliers mentioned above
(M15 and Pal5), Figure~\ref{fig:f3} shows relative distance errors $\Delta
d/d=(d_\text{phot}-d_\text{cluster})/d_\text{cluster})$ over the range in metallicity
$-2.0<\text{[Fe/H]}<-0.3$. Nine of the 11 cluster sequences considered suggest
very small systematic errors, while the distances to NGC5466 and M71 are
underestimated by 9.61\% and 17.44\%, respectively. Averaging over all
clusters, the systematic distance error is

\begin{equation}\label{eq:SEGUEsyt}
	\langle\frac{\Delta d}{d}\rangle=\begin{cases}
	   -3.28\%\pm1.78\% & \text{including M71,}\\
	   -1.86\%\pm1.19\% & \text{excluding M71.}
	   \end{cases}
\end{equation}

We found no indication in \citet{clem08} that their sequence for M71 might be
unreliable, so we will use the more conservative value of -3.28\% when
considering the effects of velocity errors on our results. This value is
indicated by the dotted line in Figure~\ref{fig:f3}. 

The photometric parallax relation of \citet{ive08} is constructed from stars
that are redder than the main-sequence turnoff. Although the authors included a
correction for age effects, and state that their formula is valid over the color
range $0.2<g-i<4.0$, we see from Figure~\ref{fig:f2} that the application of the
relationship breaks down near the turnoff, the color of which depends on both
metallicity and cluster age. Metal-poor F-type stars in the disk, for example,
have a lifetime considerably shorter than the age of the thin disk and are
already in the turnoff phase. Having turnoff stars in our sample can result in
additional systematic distance (and hence velocity) errors, which could lead to
false stream detections. Theoretical isochrones could be used to determine the
color at which stars of a given metallicity and age are in the turnoff phase.
For example, the \citet{gir04} $ugriz$ isochrones predict a turnoff color of
$g-i\approx0.36$ (0.22) for a 13.5 Gyr old stellar population with [Fe/H]$=-1.0$
($-2.0$). However, An08 have shown that the theoretical isochrones of
\citet{gir00} are not consistent with their fiducial sequences; the model colors
for the main sequence are 2-5\% too blue. Therefore, we apply a color cut
to our sample that is based on the location of the turnoff and the behavior of the
distance errors in Figure~\ref{fig:f2}, rather than on theoretical models. 

We apply a stringent color cut in order to remove turnoff stars. We choose the
color cuts depending on the metallicity of the stars as follows:

\begin{equation}\label{eq:SEGUE04}
	g-i\geq\begin{cases}
	0.55& \text{if} \text{[Fe/H]}>-1.0\\
	0.50& \text{if} -1.5<\text{[Fe/H]}\leq-1.0\\
	0.45& \text{if} -2.0<\text{[Fe/H]}\leq-1.5\\
	0.40& \text{if} \text{[Fe/H]}\leq-2.0
\end{cases}\end{equation}

These color cuts ensure that we select stars that are at least 0.05 mag redwards
of the cluster turnoffs in the corresponding metallicity bins. We do not use
stars from the most metal-rich bin for our stream search, so we do not divide
this bin further. The color cut, however, should be valid for stars up to solar
metallicity. For reference, the Sun (a G2 star) has a $g-i$ color of
$0.57\pm0.02$ \citep[as measured from about 50 solar analogs; ][]{hol06}.

To summarize, the photometric parallax relation from \citet{ive08}
performs very well in fitting the main sequences for clusters with metallicities
of $\text{[Fe/H]}\lesssim-0.3$. On average, it predicts distances that are
incorrect by $-3.28\%\pm1.78\%$. We thus adopt it to determine distances to all
the dwarfs and subdwarfs ($\log \text{g}>3.5$) in our sample. For more
metal-rich clusters ($\text{[Fe/H]}\gtrsim-0.3$) the uncertainties are generally
higher, although they can vary along the main sequence. Further
investigations with a larger number of clusters will be needed in order to
better determine the accuracy of the relation for metal-rich stars. 

\subsubsection{3D Velocities from $v_\text{rad}$, $d_\text{phot}$ and $\vec{\mu}$}

From the photometric parallax relation,
Equation~\eqref{eq:SEGUE01}-\eqref{eq:SEGUE03}, and the dereddened $r$-band
magnitudes, we calculate the distance to each star via

\begin{equation}\label{eq:SEGUE04a}
	d\text{[kpc]}=\frac{1}{100}\times10^{0.2(r-M_r)}.
\end{equation}
We calculate the statistical distance error from Gaussian error propagation:
\begin{equation}\label{eq:SEGUE05}
	\sigma_d=\frac{1}{5}d\ln10\sqrt{(\sigma_r)^2+(\sigma_{M_r})^2},
\end{equation}

\noindent where $\sigma_r$ is given for each star in our sample. The estimated
dispersion in the absolute magnitude, $\sigma_{M_{r}}$,
follows from Equations~\eqref{eq:SEGUE01}-\eqref{eq:SEGUE03}, and the listed
errors for [Fe/H] and $g-i$. The mean statistical relative distance error of our
sample is $7.58\%\pm0.01\%$. The intrinsic (systematic) scatter of
3.28\% (equation~\ref{eq:SEGUEsyt}) is less than half of this value, and can be
neglected relative to the statistical errors (see \S~\ref{sec:systdisterr}
below). Later, we restrict our selection to stars with statistical distance
errors $\frac{\sigma_d}{d}\leq15$\% and distances $d\leq 2$ kpc, so in the worst
case, when we add statistical and systematic errors, the actual distance could be
2.37 kpc instead of 2 kpc (with a total distance error of 28.28\%).

\begin{figure}[!htb]
\centering
 \includegraphics[width=\linewidth]{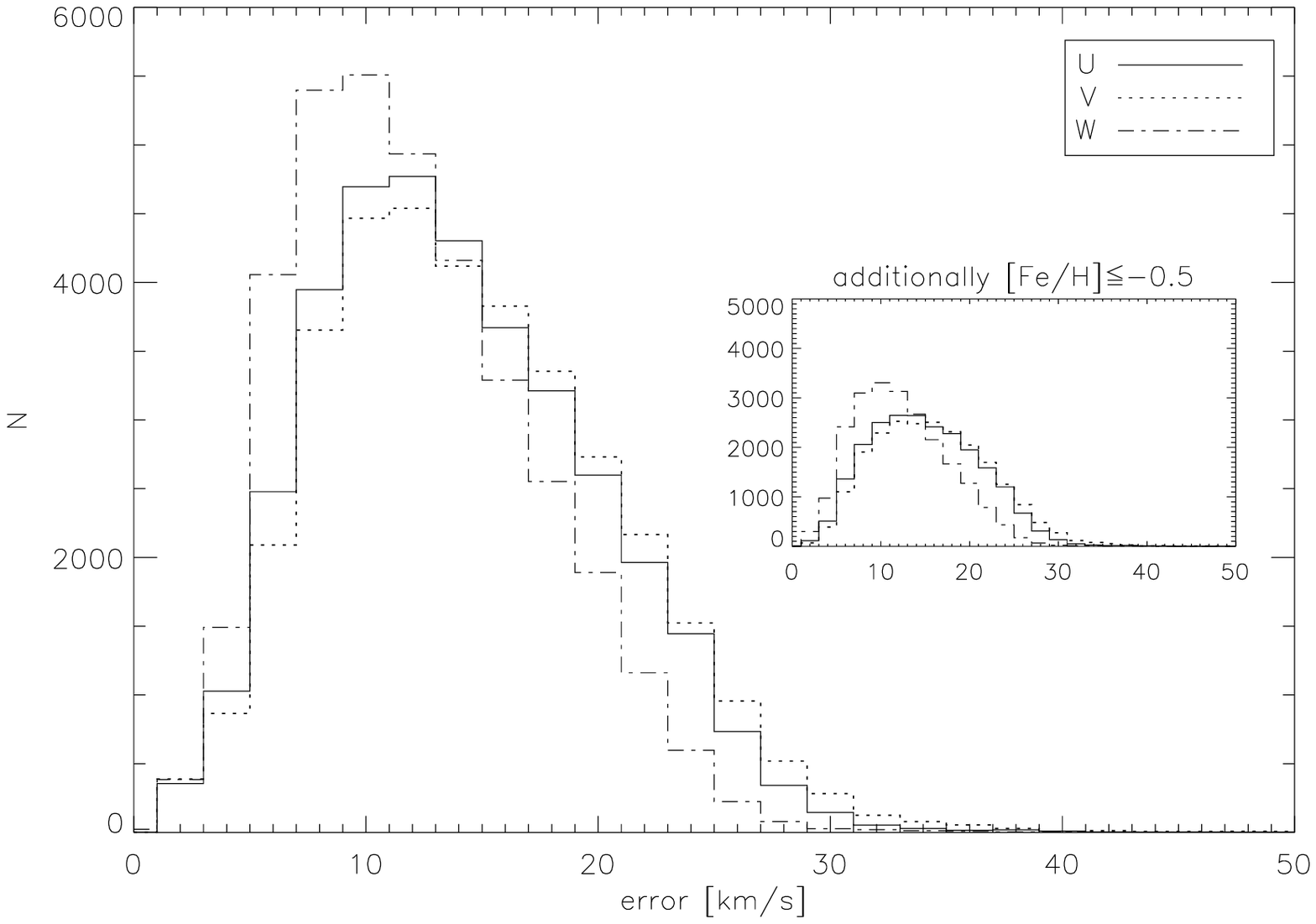}
   \caption[Distribution of velocity errors $(\sigma_U,\sigma_V,\sigma_W)$ for
   stars with SSPP flag='nnnn', $\log \text{g}\geq3.5$, $d\leq2$ kpc and
   $\frac{\sigma_d}{d}\leq15$\%] {Distribution of errors for the velocity
   components $U$ (\textit{solid}), $V$ (\textit{dotted}) and $W$
   (\textit{dash-dotted}) for all stars with flag='nnnn', $\log
   \text{g}\geq3.5$, $d\leq2$ kpc and $\frac{\sigma_d}{d}\leq15$\%. The small
   window shows the same distribution, exclusively for stars with the additional
   restriction [Fe/H]$\leq-0.5$.}
\label{fig:f4}%
\end{figure}

Figure~\ref{fig:f4} shows the statistical velocity error distribution for
37,136 stars satisfying the SSPP flag=`nnnn' (indicating no peculiarities),
$\log \text{g}\geq3.5$, $d\leq2$ kpc, $\sigma_d/d\leq15$\%, and our color cuts
(equation~\eqref{eq:SEGUE04}). Additionally, we show in the small window all
23,512 stars that also fulfill [Fe/H]$\leq -0.5$, because we cut at this
metallicity for our final sample of metal-poor stars. The velocities and their
errors have been calculated using the equations given in \citetalias[\S
2]{kle08}. 

The velocity error distribution rises to a peak around 10 km s$^{-1}$ for $W$,
and around 12--15 km s$^{-1}$ for $U$ and $V$, then falls off quickly. If we
only select metal-poor stars from the above sample, the error distribution for
$U$ and $V$ shifts to a peak around 13-15 km s$^{-1}$. For $W$ the errors
increase only slightly. This follows since the more metal-poor stars are on
average farther away. We choose to accept errors up to 35 km s$^{-1}$ for all
three velocity components. 

\subsection[A Metal-Poor Sample within 2 kpc]{A Sample of Metal-Poor Stars within 2 kpc with 6D Phase-Space Coordinates}
\label{sec:FinalSample}

In order to obtain the best available sample of stars, regarding accuracy of
distances and suitability for our stellar stream search, we only keep stars
satisfying the following criteria (the number in parentheses indicates the
number of stars that are left after each step):

\begin{itemize}
  \item $\log$ g$\geq 3.5$, in order to only select dwarfs and subdwarfs to which we can apply the photometric parallax relation (141,286)
  \item SSPP flag= `nnnn', indicating that there is are no suspected problems
with derived atmospheric parameters (118,584)
	\item distance $d\leq2$ kpc, because our search strategy requires nearby stars; also, the proper motions are more accurate for nearby stars (44,484)  
	\item relative distance errors $\sigma_d/d\leq0.15$ (44,087)
	\item total space velocity $v_{total}<600$ km s$^{-1}$, to exclude stars
with apparently false proper motion measurements or distance estimates (44,034)
	\item velocity errors smaller than 35 km s$^{-1}$ for $U$, $V$, and $W$ (43,512)
	\item $g-i\geq(g-i)_\text{TO}$, where $(g-i)_\text{TO}$ depends on the metallicity of a star according to equation~\eqref{eq:SEGUE04}, to exclude turnoff stars (35,864)
	\item we restrict ourselves to [Fe/H]$\leq-0.5$, because we concentrate
on thick-disk and halo substructure\footnote{Also, higher metallicities are
unreliable because of a calibration error that has only been fixed recently, and
was not corrected in the CAS at the time we selected our stars (Timothy Beers 2008, private communication)}(22,321)
\end{itemize}

The distance cut of 2 kpc is necessary, because our search strategy for streams
assumes a constant rotation curve in the solar neighborhood, and that we can
approximate the radial and rotational velocities by $U$ and $V$, respectively.
In addition, we gain higher accuracy in the velocities, because the proper
motions are more accurate for nearby stars. The spatial distribution of our
sample is shown in Figure~\ref{fig:SDSS-l-b} as the red dots. As can be
appreciated by inspection of this Figure, our sample is distributed over the
same region as the full DR-7 sample. 

\begin{figure}[!ht]\centering
	\subfloat[$g-i$]
		{\includegraphics[width=0.45\textwidth]{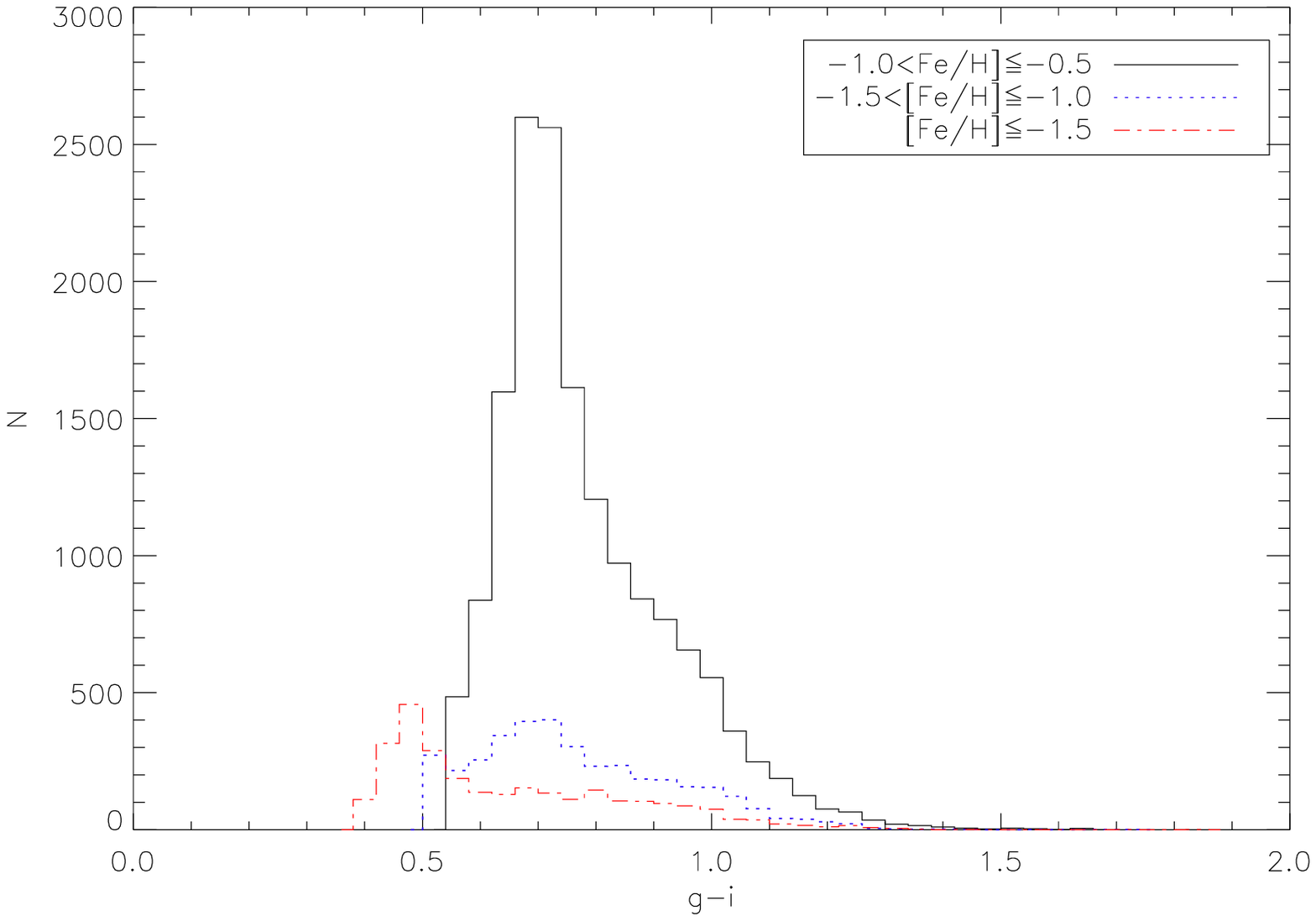}}
	\subfloat[$g-r$]
	  {\includegraphics[width=0.45\textwidth]{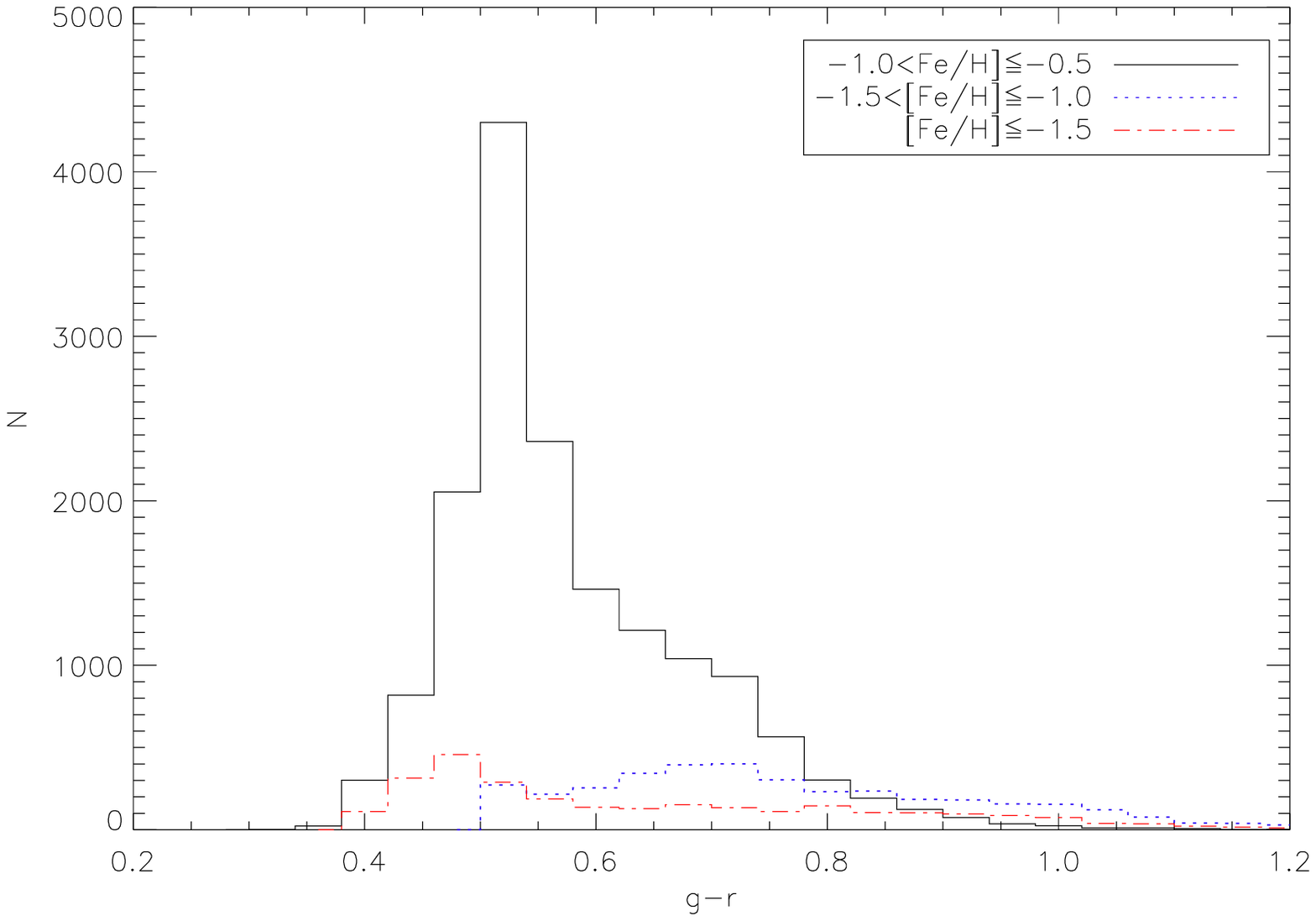}}
	\subfloat[$d$]
	  {\includegraphics[width=0.45\textwidth]{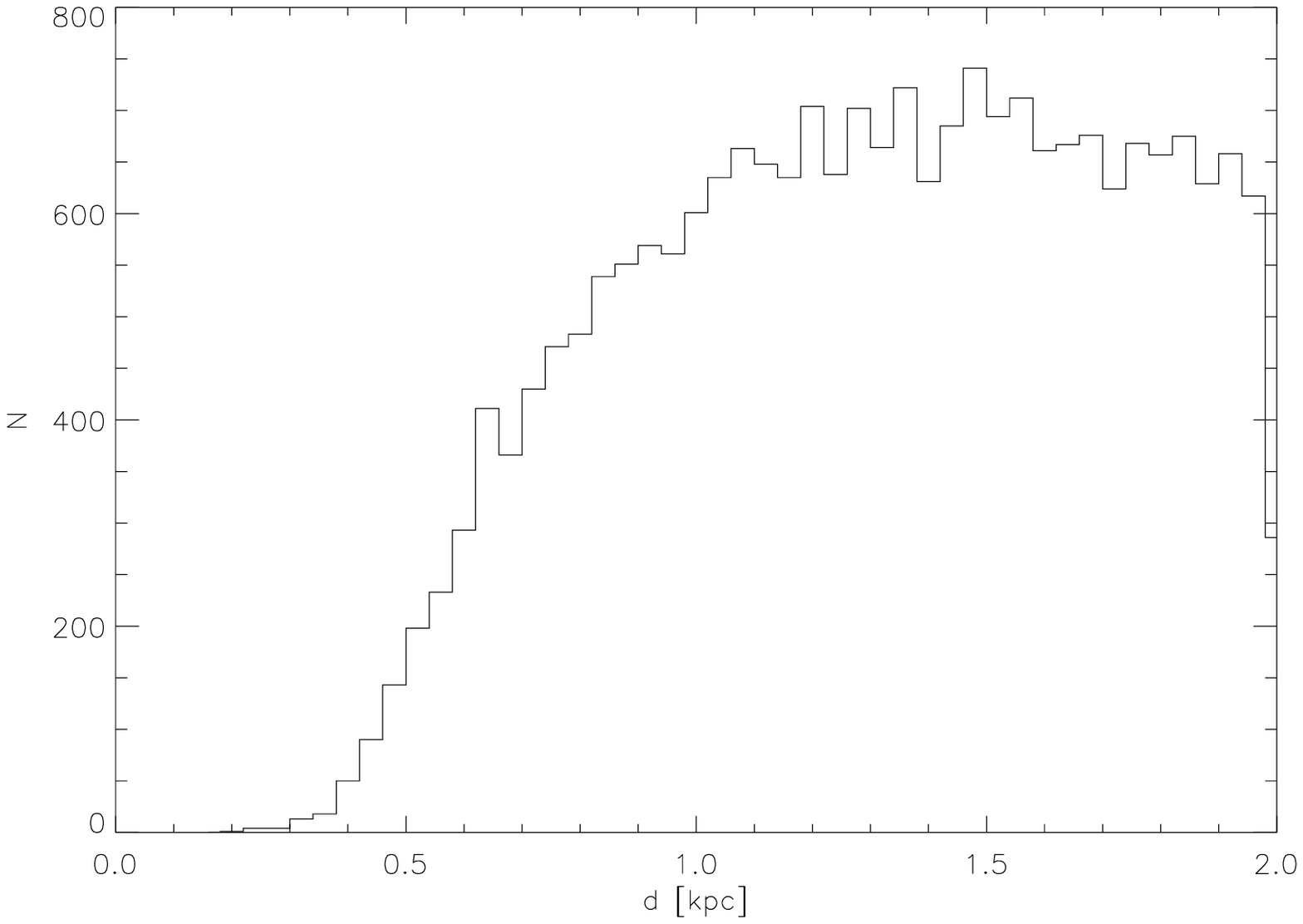}}
	\subfloat[{[Fe/H]}]
	  {\includegraphics[width=0.45\textwidth]{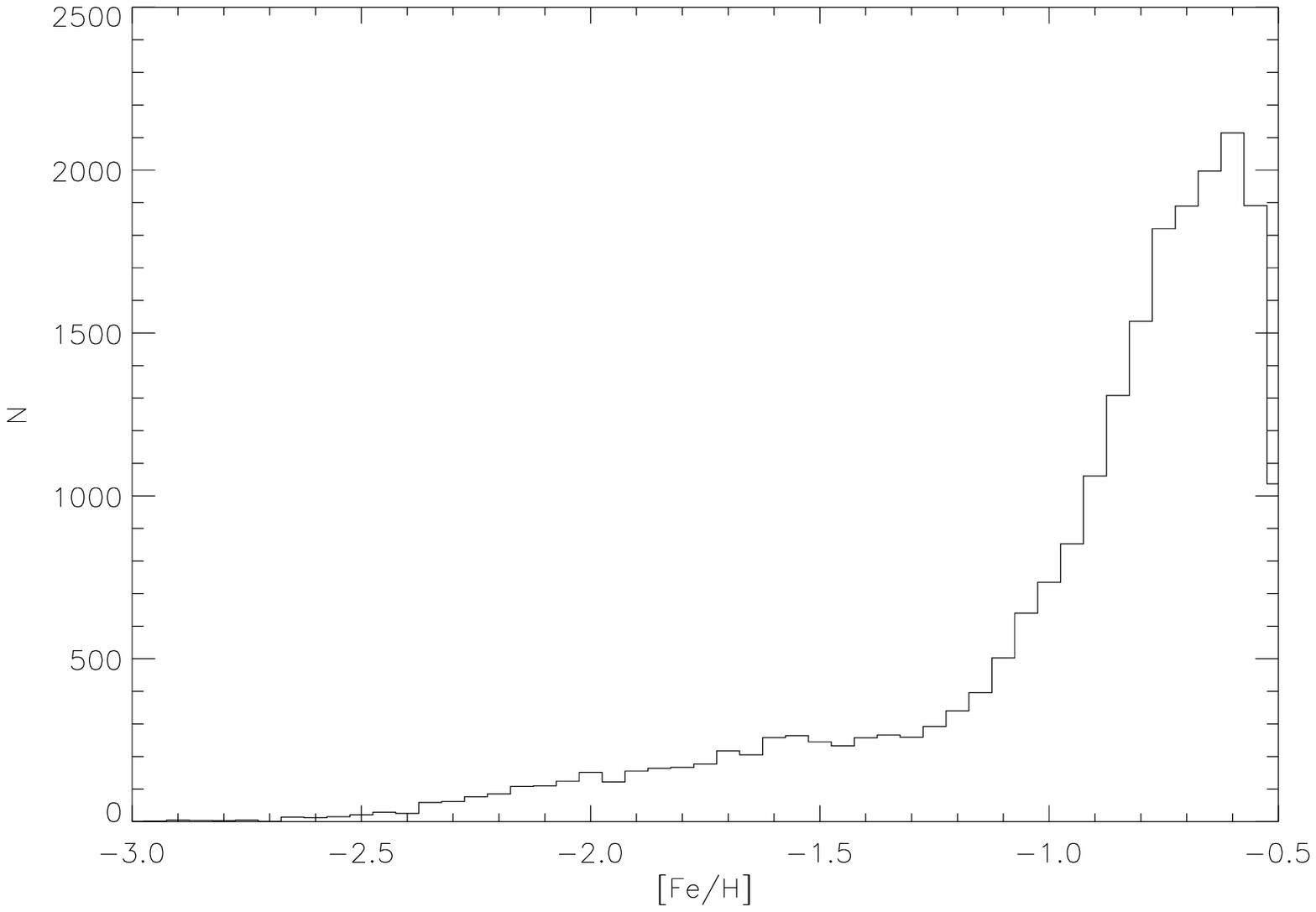}} 
        \caption[Histograms of $g-i$, $g-r$, $d$ and {[Fe/H]} for the final
        sample]{Distribution of stars from our final sample in (a) colors $g-i$,
        (b) $g-r$, (c) distance $d$, and (d) metallicity [Fe/H]. Because of the
        applied color cuts, the sample consists of G- and K-type stars.
        Panels (a) and (b) display the color distribution for three bins in [Fe/H]. The
        different panels show that our sample is dominated by stars of the
        thick-disk population.}\label{fig:coldist}
\end{figure}

Although our final sample is only 14\% of the original 154,888 stars, we have a
sample of nearby metal-poor stars of both unprecedented quantity and quality
\citep[compare to, e.g.,][]{helm99,chi00,ari06,det07}. Figure~\ref{fig:coldist}
shows the color, distance, and metallicity distribution of our final sample.
The distribution peaks at $g-i=0.7$ (the color of a G star) with a tail
extending to $g-i\approx1.3$ (K6-K7 stars). It is interesting that there remain
some possible turnoff stars in the $g-r$ distribution at $g-r\approx0.3$ for the
most metal-rich bin ($-1.0<\text{[Fe/H]}\leq-0.5$)\footnote{For the more
metal-poor bins the turnoff lies bluewards of $g-r=0.3$.}, although this is not
the case in the $g-i$ distribution. The $g-i$ color has (somewhat) better signal-to-noise properties than the $g-r$ color, except in the
region of the main-sequence turnoff \citep{ive08}. In any case, the fraction of possible turnoff stars is so small that the influence of their systematically incorrect distances on our analysis
can be neglected.

\section{Search Strategy for Streams}
\label{sec3}

To search for stellar halo streams in our sample, we effectively look for overdensities in eccentricity-,
orbital inclination-, and guiding center radius- (or, equivalently, angular momentum-) space. We adopt a method based on the
Keplerian approximation of \citet{dek76} and outlined by \citet{det07}. This
method is a generalization of the formalism outlined in \citetalias{kle08} to
identify stellar streams in RAVE data. There it was assumed that the azimuthal
velocity of a star could be approximated by $V+V_\text{LSR}$, thereby projecting
the orbits into the meridional plane. Halo stars move on random, more eccentric
orbits, and we can apply the same formalism if we group them together according
to the inclination of their orbital planes. 

We assume a spherical potential and neglect any asphericity of the dark-halo
potential and flattening of the disk potential. This is justified by the
work of \citet{chi00}, who showed that the distribution of halo stars in the
space spanned by isolating integrals of motion in an {\it aspherical} St\"{a}kel-type
potential can be closely mapped into the integrals of motion-space of a
{\it spherical} potential. Also, even for stars that move in axisymmetric flattened
potentials, $L_\perp$ is approximately conserved, and the orbits can be thought
of as planar orbits precessing slowly around the $z$-axis \citep{bt87}. As an
example we refer to \citetalias{kle08}, where it was shown that stellar thick-
and thin-disk streams detected in a projection of phase-space are also clumped
in ($L_z,L_\perp$)-space. 

In a spherical potential, a star with Cartesian velocity components $(U,V,W)$
moves in a fixed orbital plane that is inclined by an angle $\nu$ relative to
the direction towards the North Galactic Pole. The angle $\nu$ is given by

\begin{equation}\label{eq:SEGUE06}
	\nu=\arctan\bigl(\frac{V+V_\text{LSR}}{W}\bigr),
\end{equation}
and ranges from $0^\circ$ to $180^\circ$. Stars with polar angles
$\nu>180^\circ$ are treated as moving on retrograde orbits in a plane with
polar angle $\nu-180^\circ$. The azimuthal velocity of a star is
\begin{equation}\label{eq:SEGUE07}
	V_\text{az}=\sqrt{(V+V_\text{LSR})^2+W^2},
\end{equation}
so if the star is near the Sun we can approximate its total angular momentum by
\begin{equation}\label{eq:SEGUE08}
	L=R_\odot\cdot V_\text{az}=R_0\cdot V_{LSR}.
\end{equation}
For the last step we have assumed a constant rotation curve. 
Finally, we express the eccentricity, $e$, of any stellar orbit as:
\begin{equation}\label{eq:SEGUE11}
	e=\frac{1}{\sqrt{2}V_\text{LSR}}V_{\Delta\text{E}},
\end{equation}
where we have introduced the quantity
\begin{equation}\label{eq:SEGUE12}
	V_{\Delta\text{E}}=\sqrt{U^2+2(V_\text{LSR}-V_\text{az})^2}.
\end{equation}
The parameter $V_{\Delta\text{E}}$ parametrizes the difference between the energy of a
star at the guiding center of its orbit and at the solar radius,
and is a measure of its orbital
eccentricity. Also, we have shown (Burkhard Fuchs, unpublished) 
that $V_{\Delta\text{E}}$ is related to the
radial action integral, and is robust against slow changes in the gravitational
potential .\footnote{Using Dekker's \citep{dek76}
theory of Galactic orbits, the radial action integral can be expressed in the
notation of \citet{ari06} as
\begin{equation}
J_R = -\sqrt{2} \pi R_0^2\kappa_0 + \frac{\pi R_0^3\kappa_0^2}
{\sqrt{E_0 - E + \frac{1}{2} R_0^2 \kappa_0^2}} \approx 
\frac{\pi R_\odot}{2V_{LSR}} V_{\Delta E}^2 \,.
\label{eq1}
\end{equation}
} Although the approximation \eqref{eq:SEGUE11} formally breaks down for highly
eccentric orbits \citep[$e>0.5$,][]{dek76}, stars on similar orbits will still
be projected into the same region of phase space \citep{kle08a}. Looking for
``overdensities'' in $(V_\text{az},V_{\Delta\text{E}},\nu)$-space is a practical
way to find stellar streams, because we don't need to assume any expression for
the gravitational potential and any pre-history of the stream. The first
succesful application of the generalized Keplerian approximation was
by \citet{det07}, who were able to re-discover the `H99' stream, originally
found by \citet{helm99} in $(L_z,L_\perp)$-space.

\begin{figure}[!ht]\centering
	\includegraphics[width=\textwidth]{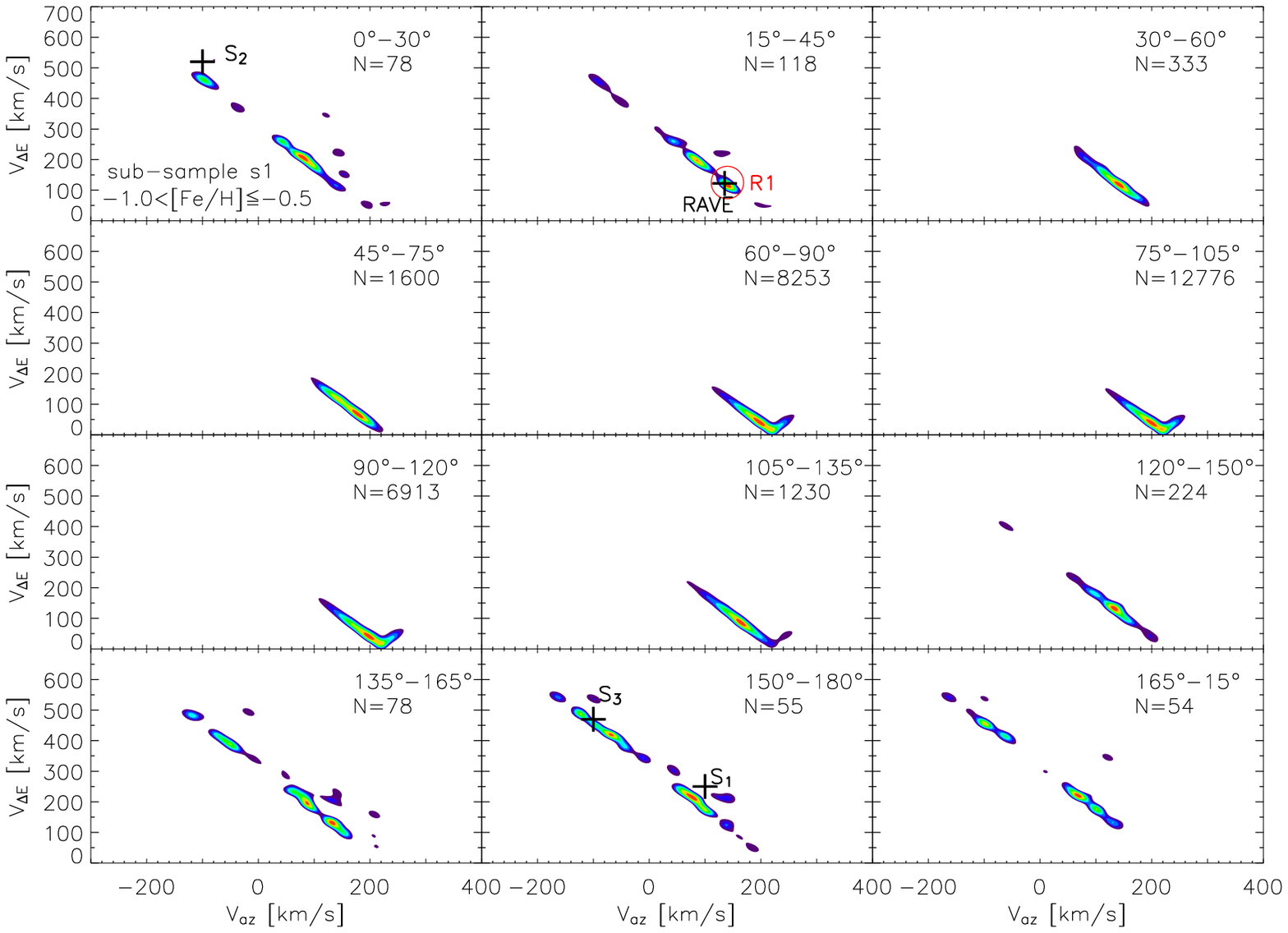}
        \caption[Wavelet transform of the $(V_\text{az},V_{\Delta\text{E}})
        $ distribution across twelve $\nu$-bins for sub-sample s1]{Wavelet
        transform of the distribution of SDSS stars from the metallicity sub-sample
        s1 (see text for definition) in $(V_\text{az},
        V_{\Delta\text{E}})$-space, shown in bins of different orbital
        polar angle. The contours in each $\nu$-bin range from 10\% to 100\% of
        the maximum value of the wavelet transform, and are color coded
        accordingly from purple to red (see Fig.~\ref{fig:erroreffect} for the colorbar). We have marked the positions of 
        known stellar streams given in Table~\ref{tab:stcoords} with
        thick crosses and circled in red overdensities.  See
        \S~\ref{sec7} for discussion.}\label{fig:wl-s1}
\end{figure}

\begin{figure}[!ht]\centering
	\includegraphics[width=\textwidth]{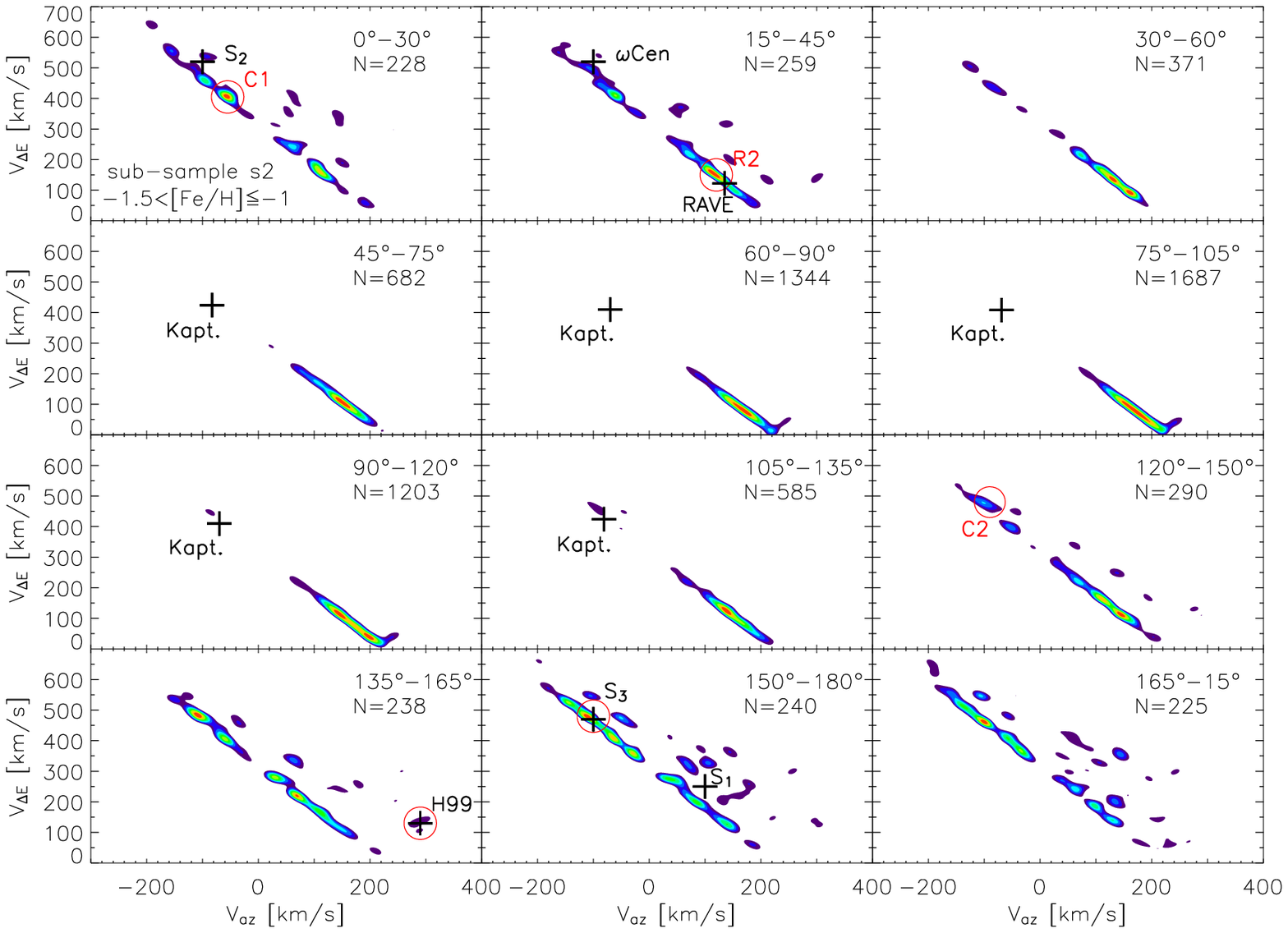}
        \caption[Wavelet transform of the $(V_\text{az},V_{\Delta\text{E}})
        $ distribution across twelve $\nu$-bins for sub-sample s2]{Same as
        Figure~\ref{fig:wl-s1}, but now for stars in the metallicity sub-sample
        s2.}\label{fig:wl-s2}
\end{figure}

\begin{figure}[!ht]\centering
	\includegraphics[width=\textwidth]{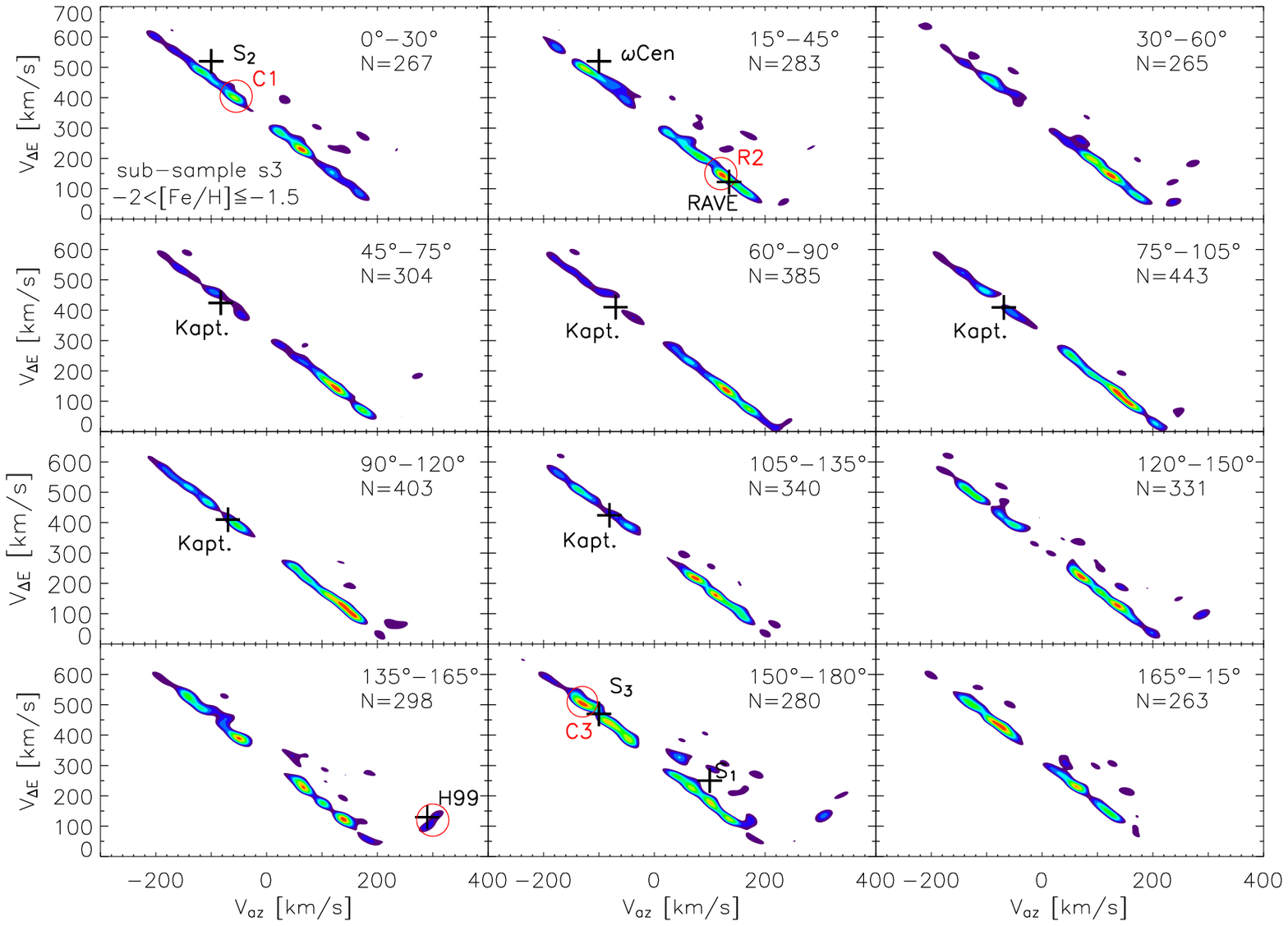}
        \caption[Wavelet transform of the $(V_\text{az},V_{\Delta\text{E}})
        $ distribution across twelve $\nu$-bins for sub-sample s3]{Same as
        Figure~\ref{fig:wl-s1}, but now for stars in the metallicity sub-sample
        s3.}\label{fig:wl-s3}
\end{figure}

\begin{figure}[!ht]\centering
	\includegraphics[width=\textwidth]{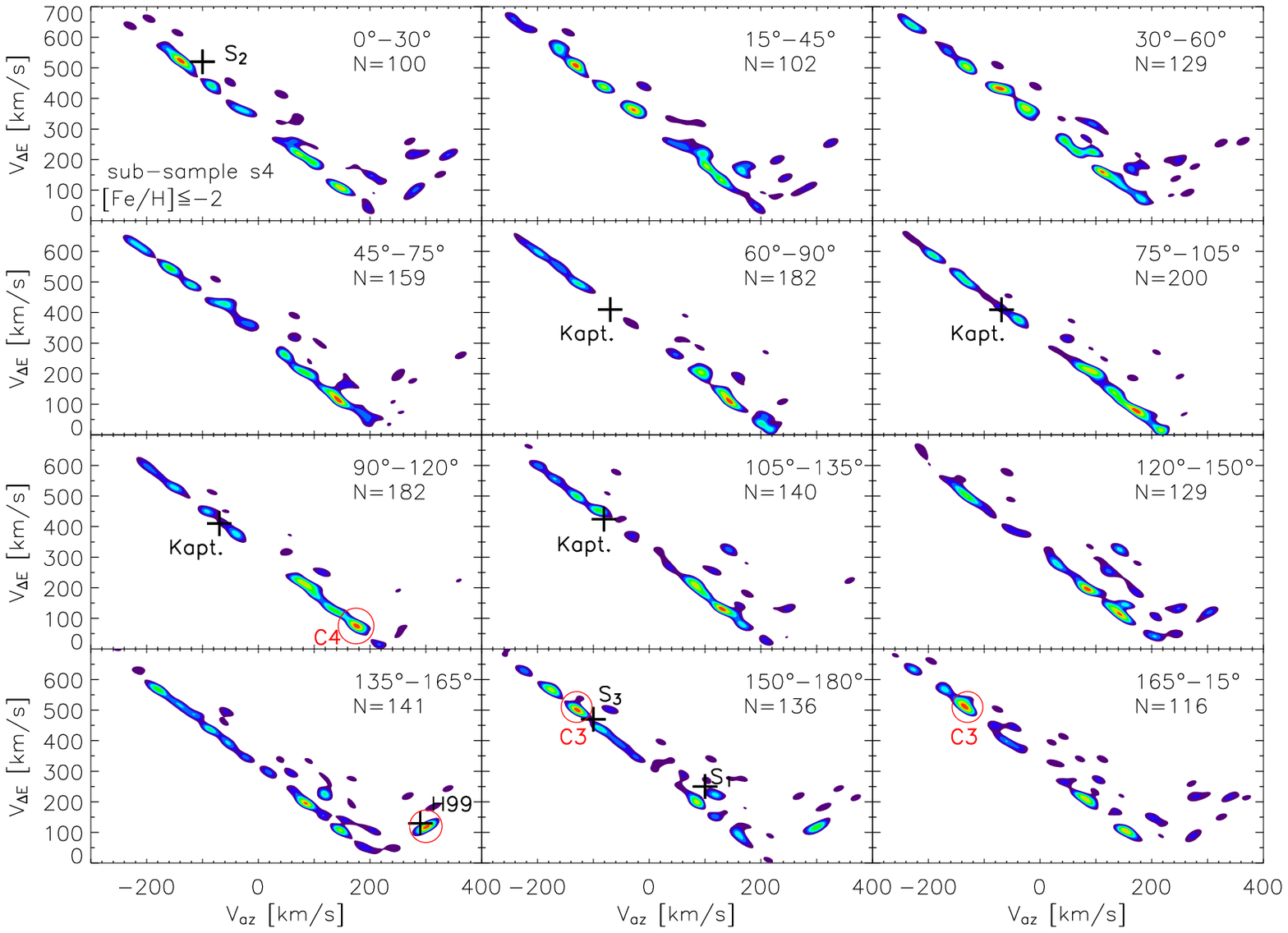}
        \caption[Wavelet transform of the $(V_\text{az},V_{\Delta\text{E}})
        $ distribution across twelve $\nu$-bins for sub-sample s4]{Same as
        Figure~\ref{fig:wl-s1}, but now for stars in the metallicity sub-sample
        s4.}\label{fig:wl-s4}
\end{figure}

We use metallicities [Fe/H] to discriminate between stellar populations with
different origins. We divide our sample into four sub-samples (hereafter s1,s2,
s3,s4) with decreasing metallicity:

\begin{itemize}\label{item:subsamples}
	\item s1: $-1.0<\text{[Fe/H]}\leq-0.5$ (15856)
	\item s2: $-1.5<\text{[Fe/H]}\leq-1.0$ (3676)
	\item s3: $-2.0<\text{[Fe/H]}\leq-1.5$ (1931)
	\item s4: $\text{[Fe/H]}\leq-2.0$      (858)
\end{itemize}
The number in parenthesis is the number of stars in the corresponding
sub-sample. We note that the typical metallicity
estimates are uncertain to (at best) $\sim$0.25 dex, making adjacent sub-samples
not entirely independent. For each [Fe/H]-bin, we collect the stars with similar orbital
polar angles in small $\nu$-slices. We bin the polar angles into
$30^\circ$-wide bins that overlap by $15^\circ$, thereby reducing bin-boundary
effects on the results. We conduct the search for stellar streams within each
$\nu$-slice in the space spanned by angular momentum and eccentricity, or
$(V_\text{az},V_{\Delta\text{E}})$. To amplify the overdensities we use a
wavelet transform technique with a skewed Mexican-hat-shaped analyzing wavelet
(see \citetalias{kle08} for more details). We set the scale parameter of the
wavelet to $a=12$ km s$^{-1}$, comparable to the velocity errors, set the
elongation parameter to $q=\sqrt{3}$, and employ cells in $(V_\text{az},
V_{\Delta\text{E}})$-space of 3 km s$^{-1}$ width on each side. The resulting
contours of the wavelet transform are shown in
Figures~\ref{fig:wl-s1}-~\ref{fig:wl-s4}. 


\section{Placing Solar Neighborhood Streams in the $(V_\text{az}$,$V_{\Delta\text{E}})$-Plane}
\label{sec4}

As a first step in the analysis of the SDSS data, we explore whether we can find
evidence in these data for the known streams believed to be of
a tidal origin. We have taken the velocities of known halo streams from
the literature and evaluated their velocities $V_\text{az}$,
$V_{\Delta\text{E}}$, and orbital inclinations, $\nu$. The results are given in
Table~\ref{tab:stcoords} and marked with thick crosses in
Figures~\ref{fig:wl-s1}-~\ref{fig:wl-s4}. The feature marked `S$_2$', 
found by \citet{det07} as an overdensity of stars in $(V_\text{az}$,
$V_{\Delta\text{E}}$,$\nu$)-space, has almost exactly the same properties as the
`$\omega$ Cen'-stream; in fact, the signal of `S$_2$' in Figure~3 of
\citet{det07} is extended across the $\nu$-range from 0 to $\sim20^\circ$.
Therefore, it is likely that `S$_2$' is in fact the `$\omega$ Cen'-stream, and we
only label `$\omega$ Cen' in the plots. 

\begin{deluxetable}{c|c|c|c|c|c|c|c|c}
\tablecaption{Velocities and Derived Effective Integrals of Motion for Known Solar
Neighborhood Streams\label{tab:stcoords}}
\tabletypesize{\scriptsize}
\tablehead{
\colhead{} & \colhead{Kapteyn\tablenotemark{a}} &  \colhead{H99\tablenotemark{b}} & \colhead{$\omega$ Cen\tablenotemark{c}} & \colhead{RHLS\tablenotemark{d}} &   \colhead{S$_1$\tablenotemark{e}} & \colhead{S$_2$\tablenotemark{e}} & \colhead{S$_3$\tablenotemark{e}} & \colhead{RAVE\tablenotemark{f}}
} 
\startdata
			$\langle\lvert U\rvert\rangle$ & 63 & 84 & \nodata  & 294 & \nodata & \nodata & \nodata & 24\\
			\hline $\sigma_{\lvert U\rvert}$ & 54 & 65 & \nodata  & 6  & \nodata  & \nodata  & \nodata  & 15\\
			\hline $\langle V\rangle+V_{LSR}$ & -69& 130 & \nodata  & 99 & \nodata & \nodata & \nodata & 61\\
			\hline $\sigma_V$ & 6 & 22 & \nodata  & 25 & \nodata & \nodata& \nodata & 5\\
			\hline $\langle W\rangle$ & -16 & -240 & \nodata & 239 & \nodata & \nodata & \nodata & 121\\
			\hline $\sigma_W$ & 67 & 24 &  \nodata & 24 & \nodata & \nodata & \nodata & 30\\
			\hline $\langle V_\text{az}\rangle$ & -71 & 273 & -100 & 259 & 100 & -100 & -100 & 135 \\
			\hline $\langle V_{\Delta\text{E}}\rangle$ & 416 & 128 & 520 & 302 & 250 & 520 & 470 & 122\\
			\hline $\langle\nu\rangle$ & 77 & 152 & 25 & 22 & 165 & 6 & 170 & 27\\
			\hline expected in sub-sample & s1,s2,s3 & s2,s3,s4 & s2,s3 & s3,s4  & unknown & unknown & unknown & unknown
			\\
\enddata
		
\tablenotetext{a}{\citet{egg96}}
\tablenotetext{b}{\citet{helm99,kep07}}
\tablenotetext{c}{\citet{din02,brook04a}}
\tablenotetext{d}{\citet{refi05}}
\tablenotetext{e}{\citet{det07}}
\tablenotetext{f}{\citetalias{kle08}}

\tablecomments{All velocities are given in km s$^{-1}$; the
$\nu$-angles in $^\circ$. Note that `S$_2$' is probably the `$\omega Cen$'
stream.} 
\end{deluxetable}

We find strong evidence for the `H99', `S$_3$', and `RAVE' streams, which we 
discuss more in \S~\ref{sec7}, but little
or only minor hints for the existence of the other known streams.
\clearpage

\clearpage
\section{The Effects of Systematic Distance Errors and Unresolved Binaries}
\label{sec:systdisterr}
 
\begin{figure}[!ht]\centering
  \subfloat[without additional distance errors]{
	\includegraphics[width=0.45\textwidth]{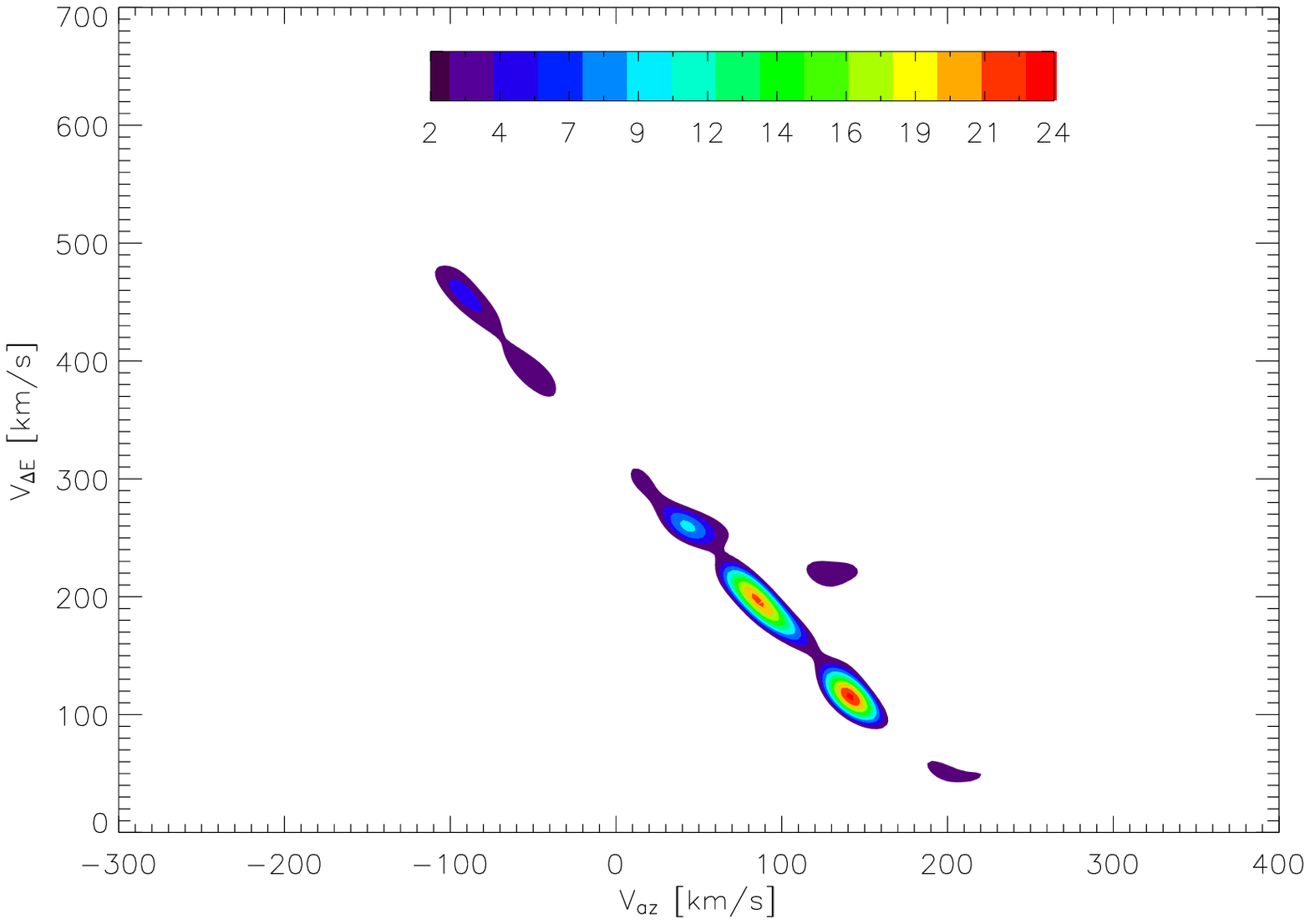}\label{systa}}
	\subfloat[with additional -10\% distance errors]{
	\includegraphics[width=0.45\textwidth]{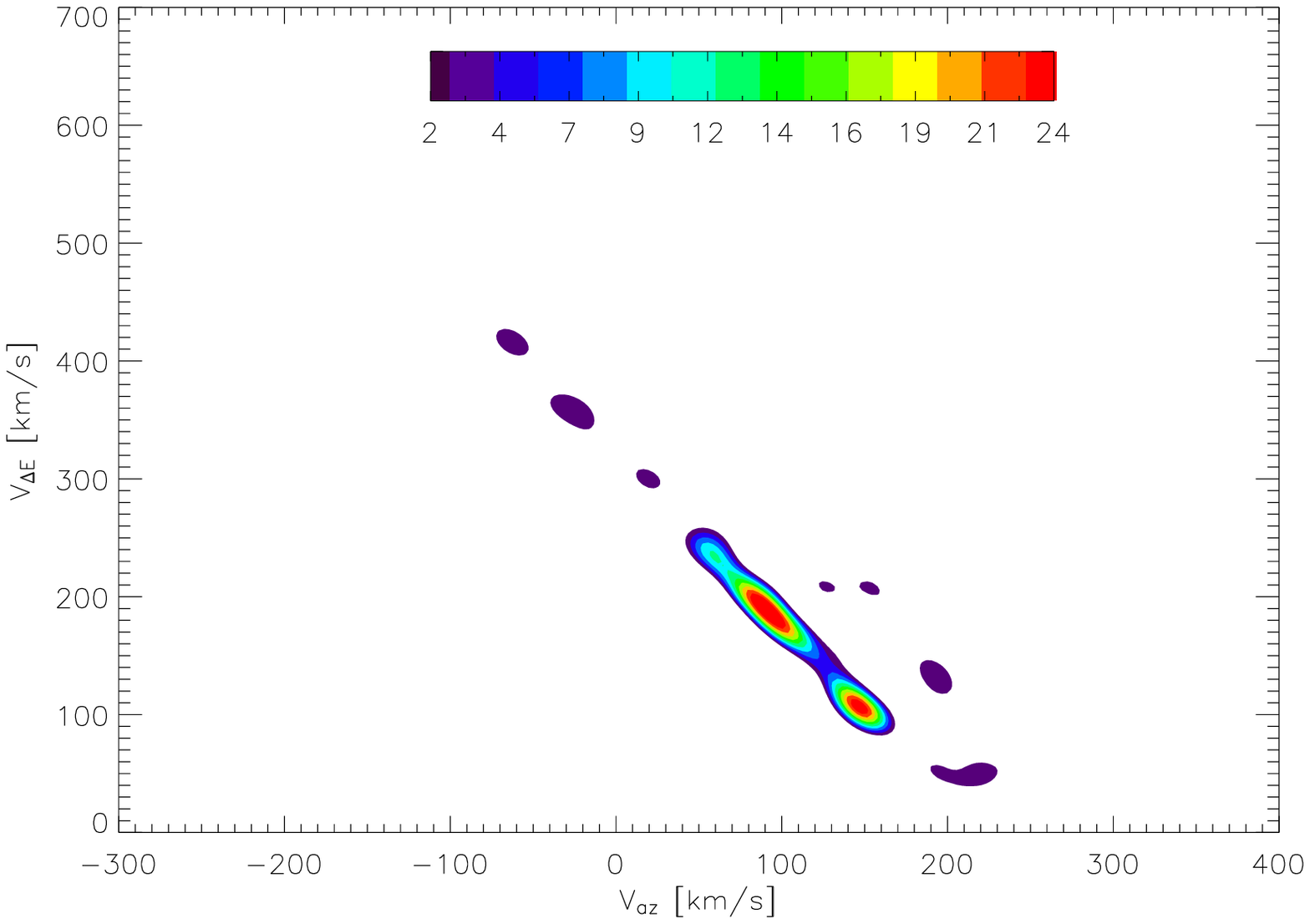}\label{systb}}
        \caption[(a) Effect of the systematic distance error on the $(V_\text{az}$,
        $V_{\Delta\text{E}})$ distribution]{Contours of the wavelet transform 
        for the distribution of stars from sub-sample s1 that lie in the $\nu$-range
        $15^\circ$-$45^\circ$ (b) The same distribution, but
        after adding a systematic distance error of -10\% to the data. Note
        how the overdensities change shape and relative ``height", but fairly keep
        their position.\label{fig:erroreffect} }
\end{figure}

To test for the possible effects of systematic distance errors on our results,
we have added the -3.28\% systematic distance error that we found through
comparison of the photometric parallax relation to cluster fiducial sequences in
\S~\ref{sec:ComparisonWithFiducials} to each star, then re-calculated
their velocities and the wavelet transform for a subset of stars. As expected, the effect of the additional -3.28\% errors 
is to slightly change the shape and relative ``height" of some features, but the location of
the overdensities does not change much. However, we also want to test if a much higher systematic error -- say, -10\% -- would effect 
the location of the overdensities. While we have shown in \S~\ref{sec:ComparisonWithFiducials} 
that a -10\% error is very unlikely at the low-metallicty end of the photometric parallax relation, 
the error at the high-metallicity end ([Fe/H]$\gtrsim-1.0$) is much less constrained and a possibility for systematic
errors on that level exists. This in turn may cause an overdensity that
has a range of metallicities to be detected at varying positions in the
($V_\text{az}$,$V_{\Delta\text{E}}$)-plane in different metallicity sub-samples, and be
misinterpreted as multiple, adjacent, streams.  As an example,
Figure~\ref{fig:erroreffect} shows contours of the wavelet transform for the
distribution of stars from sub-sample s1 in the $\nu$-slice
$15^\circ$-$45^\circ$ \subref{systa} without and \subref{systb} 
with additional -10\% distance errors. While the features remain close to their locations
in ($V_\text{az}$,$V_{\Delta\text{E}}$), they change their shape and relative ``height". We have investigated other sub-samples and $\nu$-slices with adding similar distance errors (from 7\% to 10\%) and found that the change in relative height of the overdensities is more prominent for stars on retrograde orbits. However, such stars are more represented in the lower metallicity bins, where we expect the systematic distance errors to be less severe. Given that the positions of the peaks roughly remain where they are in all examples we investigated, we can assume that the location of stellar streams in the ($V_\text{az}$,$V_{\Delta\text{E}}$)-plane is not very sensitive to the expected systematic distance errors. 

However, problems can arise when we want to distinguish 
adjacent streams that span more than one metallicity bin. Because the systematic errors can change with metallicity, the signal of a single stream can, too, and in this way produce apparently distinct, but adjacent peaks.

Unresolved binary stars may affect our analysis through their effect on the photometric parallax relation and hence the distance determination. The number of multiple stellar systems expected in our sample of mainly G and K stars might be taken to be around 57\% (probably lower) according to recent determinations \citep[and references therein]{lada06}.\footnote{We note, however, that in our case of metal-poor stars the expected fraction of multiple stars drops further, because these stars tend to be older, hence the systems had more time to disperse.} Unresolved multiple stars taken as single stars have too high observed luminosities and hence their distances are underestimated. Another effect arises, if the multiple stellar systems consisits of stars of different masses, which leads to a shift in colors. To get a feeling of how the distances change when we mistake a binary system as a single star, we calculate by how much the apparent magnitude, $m$, changes. In the case of equal-mass binaries the observed flux, $F$, would be only half as high for a single star.
\begin{equation}
	m=-2.5\log{\frac{F}{F_0}}
\end{equation}
is the definition of the apparent magnitude, where $F_0$ is some constant flux which gives $m=0$. It follows that the apparent magnitude that would be assigned to only the single star from the binary system, $m'$, is given by
\begin{equation}
	m'=-2.5\log{\frac{0.5F}{F_0}}\simeq m+0.75.
\end{equation}
Together with equation~\eqref{eq:SEGUE04a}, it follows that in the worst case (equal-mass binaries) the distances for part of our sample are underestimated by 29\%. This is a systematic effect, and results in a change of the shape and relative ``height" of the overdensities as discussed above. However, this effect might not be so severe, because (a) the change in distances has the opposite sign to the expected systematic error of the photometric parallax relation, (b) it affects approximately only half of our sample, and (c) most of nearby G-dwarf binary systems possess mass ratios that peak at much less than 1.0 \citep{duq91}.  In addition to the systematic underestimation of distances, the radial velocities for multiple systems are scattered around the value of the dominant component in a statistical way due to the additional doppler shift in the spectrum of the bound system. Taken together, these effects are hard to describe quantitatively, but they certainly only dilute the signatures of real streams. Hence this works to make the numbers we identify as stellar streams a lower limit of what might exist.

\section{Estimating the Significance of Overdensities}
\label{sec6}

It is clear that overdensities in $(V_\text{az},V_{\Delta\text{E}})$ can and
will emerge through Poisson noise, especially in regions that are sparsely
sampled by the data. As described in \citetalias[\S~4.2]{kle08}, we can
address this problem by performing Monte Carlo simulations with stars randomly
drawn from a smooth distribution. From these we can build the residuals of the
wavelet transforms of each individual simulation against the mean value for all
wavelet transforms, which represents a smooth distribution over our search
space. For each cell in $(V_\text{az},V_{\Delta\text{E}})$, we can calculate the
variance and use it to obtain a significance map of the
overdensities\footnote{Artificially high significance levels can emerge if we
divide the residuals by a standard deviation which is less than 1. For this
reason, we set the variance to 1 in each cell where it is below this level.}.

For the smooth halo and thick-disk components, we adopt the results of
\citet{chi00}, who characterized the halo through a mean rotational velocity at
$\langle \theta \rangle\approx30$ km s$^{-1}$, with a radially elongated velocity
ellipsoid $(\sigma_U,\sigma_V,\sigma_W)=(141\pm11,106\pm9,94\pm8)$ km s$^{-1}$.
We further use $\langle \theta \rangle=190$ km s$^{-1}$ and $(\sigma_U,\sigma_V,
\sigma_W)=(46\pm4,50\pm4,35\pm3)$ for the thick disk, and adopt their estimated
fraction of thick-disk stars in our sub-sample s1 as 80\%, in s2 as 30\% and in
s3 as 10\%. We are aware that the azimuthal drift for the thick-disk stars we
adopt is at the lower end of values given in the literature; for example, \citet{sou03} find
$\langle \theta \rangle=159\pm5$ km s$^{-1}$ from spectoscopic and kinematical
analysis of nearly 400, mostly clump-giant, stars. However, their sample was
limited to stars with abundances [Fe/H]$>-0.65$ and distances $d\lesssim800$ pc,
so the \citet{chi00} data more closely resemble our own data.  It should also
be noted, for the purpose of this exercise, that we have not explicitly included the
possible presence of stars from an outer-halo component. As Carollo et al. (2007)
have argued, such stars only begin to dominate 15-20 kpc from the Galactic
center, and are not likely to comprise a major component in our solar
neighborhood sample.

\begin{figure}[!ht]\centering
	\includegraphics[width=\textwidth]{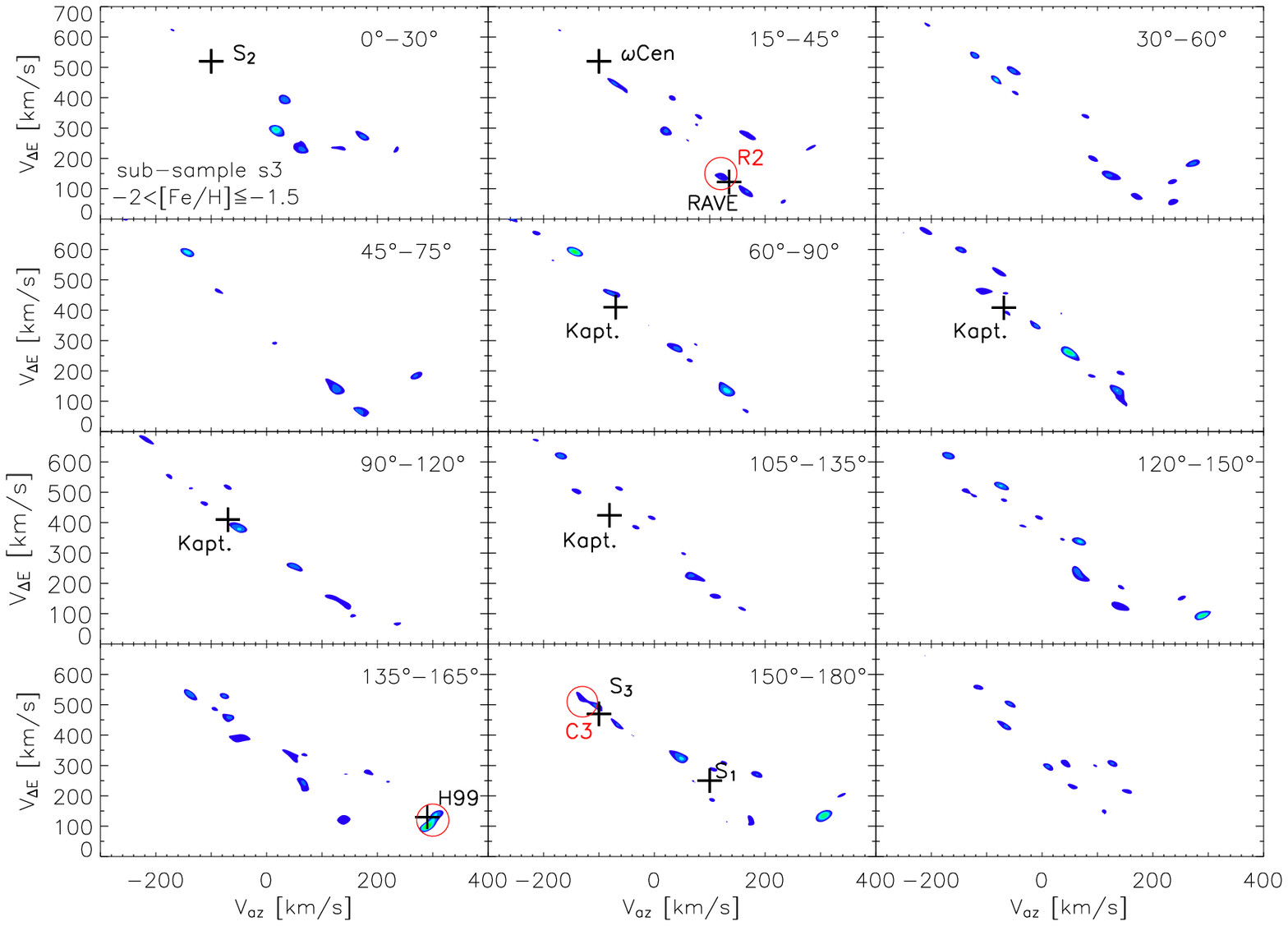}
	\caption[Significance map for sub-sample s3]{Significance map of the overdensities from Figure~\ref{fig:wl-s3}. Only areas with a $\sigma\geq2$ are shown. The contours range from 2 (blue) to 10 (red). Note the abundance of stars on disk-like orbits. The crosses mark the expected position of already known stellar streams. Kapteyn's stream appears more than once, because it is well-defined in $(U,V)$, but not $W$, so it spans a broader range of orbital inclinations (because $\nu=\arctan\frac{V}{W}$).}\label{fig:sig-s3}
\end{figure}

To each velocity drawn from the smooth kinematic models (with Gaussian velocity
ellipsoids) we assigned as an additional velocity error the mean velocity error
of our data, that is, $\bigl(\langle\Delta U\rangle$,$\langle\Delta V\rangle$,
$\langle\Delta W\rangle\bigr)$=$(14.9,15.7,12.2)$ km s$^{-1}$. For each
metallicity sub-sample we built 30 Monte Carlo realizations with (by
construction) smooth velocity distributions, consisting of the same number of
stars as the sub-sample. The Monte Carlo samples were then analyzed in the same
way as the real data, that is, through collecting the stars in different
$\nu$-bins and performing the wavelet analyses in these bins. Because the number
of stars in each $\nu$-bin is small, we expect a considerable amount of
shot-noise. Also, because the velocity dispersions are very large, small
deviations from our choice of the velocity ellipsoid would probably result in
large changes of the significance levels. In addition, a stellar stream populating
a certain part of phase-space with stars will increases the level of
Poissonian fluctuations, which are proportional to $\sqrt{N}$. When we divide
the residuals between our smooth model and these fluctuations by the expected sigma for the smooth model, it may lead to the
appearance of apparently highly significant multiple, adjacent, peaks
(while in fact they are a part of the same structure). We therefore treat the significances with care and keep in mind
that one single stream can produce multiple close features. Figure~\ref{fig:sig-s3} shows, as an example,
the significance map for sub-sample s3, where we have only displayed areas with
$\sigma\geq2$. We inspected the significance maps for the other
sub-samples as well, and derived significance levels for all putative stellar
streams \citep[for more details see][]{kle08a}. Individual results are
discussed in the next section.

\section{The Results: Stellar Halo Streams}
\label{sec7}

Here we explore the best ``slicing'' in orbital parameter- and metallicity-space
by using ($V_\text{az}$,$V_{\Delta\text{E}}$,\\$\nu$,[Fe/H]). Metallicity is an
additional constraint to distinguish tidal and dynamical streams, or different
streams that occupy the same region in $(V_\text{az},V_{\Delta\text{E}},\nu)
$-space. Tidal debris still carries the chemical information of its progenitor,
while dynamical streams are composed of stars that lack a common origin. In
addition, the latter will contribute to our sample to a lesser extent as the
metallicity decreases, because dynamical streams are dominated by disk stars. In
the following we will discuss all the ``phase-space overdensities'' that we
either identified as previously known streams or likely candidates for new
stellar halo streams. The preconditions for selecting these stream candidates
included a sufficiently large number of stars ($N$>15), a significance level
$\sigma\geq2$, and a metallicity distribution consistent with a tidal origin.
Table~\ref{tab:t2} summarizes the main properties of the identified streams.
Lists of the putative stream members, their stellar parameters, and their
derived distances and kinematic parameters, are provided in the Appendix.
We first consider any smooth components
in the phase-space distribution, then concentrate on putative halo stars on
non-disk-like orbits.

\begin{deluxetable}{c|c|c|c|c|c|c|c|c}
\tabletypesize{\scriptsize}
\tablewidth{0pt}
\tablecaption{Main Characteristics of the Stellar Streams Considered Real\label{tab:t2}} 
			\tablehead{\colhead{Stream} & \colhead{References} & \colhead{V$_\text{az}$} & \colhead{V$_{\Delta\text{E}}$} & \colhead{$\nu$} & \colhead{$\sigma$} & \colhead{$N$} & \colhead{$\langle\text{[Fe/H]}\rangle$} & \colhead{$\sigma_\text{[Fe/H]}$}\\
\colhead{} & \colhead{} & \colhead{(km s$^{-1}$)} & \colhead{(km s$^{-1}$)} & \colhead{($^\circ$)} & \colhead{} & \colhead{} & \colhead{} & \colhead{}}
	\startdata
			H99 & \citet{helm99} & 300 & 120 & 150 & 12.0 & 21 & -1.8 & 0.4\\
			RAVE & \citet{kle08} & 120 & 150 & 30 & 3.0 & 19 & -1.4 & 0.3\\
			S$_3$ & \citet{det07} & -100 & 470 & 155 & 4.5 & 33 & -1.6 & 0.4\\
			C1 & new & -60 & 410 & 15 & 2.9 & 32 & -1.5 & 0.2\\
			C2 & new & -100 & 470 & 135 & 3.4 & 53 & -1.6 & 0.4\\
			C3 & new & -130 & 510 & 170 & 3.7 & 44 & -1.7 & 0.4\\
			C4 & new & 175 & 75 & 100 & 4.8 & 20 & -2.3 & 0.3\\
		\enddata
 \tablecomments{The parameter $\sigma$ denotes the significance of the stream obtained as
described in \S~\ref{sec6}, $N$ is the number of putative stream
members, $\langle\text{[Fe/H]}\rangle$ is the mean metallicity of the stream, and
$\sigma_\text{[Fe/H]}$ the standard deviation.}
\end{deluxetable}

\clearpage
 
\subsection{The Smooth Thick-Disk Component}

We now discuss the phase-space distribution of the stars, starting with the
highest metallicity bin, s1 (Figure~\ref{fig:wl-s1}). Stars in this metallicity
range, between [Fe/H]$=-0.5$ and $-1.0$, should be dominated by members of the
thick disk; this is clearly discernible as the smooth distribution of stars with
polar angles around $90^\circ$ that rotate with $\langle
V_\text{az}\rangle\approx200$ km s$^{-1}$ around the Galactic center. In the
significance map we have also found a hint of the `Hercules' stream at
$(V_\text{az},V_{\Delta\text{E}})\approx(169,73)$ km s$^{-1}$. 

For more highly inclined orbits, the distribution of stars peaks at lower
azimuthal velocities (see Figure~\ref{fig:wl-s1}), indicating that the fraction
of halo stars increases. In addition, at angles of
$\lvert\nu-90^\circ\rvert\gtrsim45^\circ$, overdensities of stars on retrograde
orbits begin to emerge. There is no more smooth stellar component on such
orbits. At these orbital inclinations, a few stars are sufficient to create an
artificially high signal, because the variances of the residuals between the
single Monte Carlo realizations and their smooth superposition are nominally
very small. Therefore, we question whether such overdensities in s1 are real, unless
they also show up in the other metallicity sub-samples.  

The amount of substructure increases for the sub-sample s2, which consists of
stars in the metallicity range $-1.5<\text{[Fe/H]}\leq-1.0$ (Figure~\ref{fig:wl-s2}). There
are still a large numbers of stars on disk-like orbits, with 46\% of the stars
having orbital inclinations between $75^\circ$ and $105^\circ$. The mean
rotational lag with respect to the LSR increases further. According to
\citet{chi00}, $\langle \theta \rangle$ decreases linearly with [Fe/H] for
[Fe/H]$\gtrsim-1.7$ and stays approximately constant below [Fe/H]$=-1.7$. 

It is hard to make out substructure among the stars on disk-like orbits, but we
think we see a hint of the stream `AF06' detected by \citet{ari06} and
\citet{helm06}. Its stars have orbital inclinations ranging from slightly below
to slightly above the Galactic plane, depending on whether we take stars with
positive or negative $W$ velocities from the list of member stars given in
\citet{ari06}. We detect a signal of this stream at $(V_\text{az},
V_{\Delta\text{E}})\approx(140,110)$ km s$^{-1}$ in the $\nu$-slice
$90^\circ$-$120^\circ$, maybe even extending towards $105^\circ$-$135^\circ$. 

In the sub-sample s3, at metallicities in the range $-2.0<\text{[Fe/H]}\leq-1.5$
(Figure~\ref{fig:wl-s3}), the fraction of halo stars dominates over thick-disk
stars. In the regions of the dynamical streams `Hercules' and `AF06' we detect
signals that are significant above $\sigma\geq2$. These signals remain even for
the sub-sample s4 (with [Fe/H] $< -2.0$), where the fraction of canonical
thick-disk stars should be negligible, showing that still some (metal-weak)
thick-disk stars are present. 

\subsection{Confirming the Discovery of the RAVE DR-1 Stream} 

We now consider the location of the stream discovered in the
RAVE DR-1 data discussed in \citetalias{kle08}. This stream is centered at a mean of
$(V_\text{az},V_{\Delta \text{E}})\approx(135,122)$ km s$^{-1}$ and
$\nu\approx30^\circ$.\footnote{In the analysis of the RAVE stars,
\citetalias{kle08} projected their azimuthal motions onto the Galctic plane by
adopting a cylindrical coordinate system and setting $V_\text{az}=V$,
$V_{\Delta\text{E}}=\sqrt{U^2+2(V_\text{az}-V_\text{LSR})^2}$. This
worked because the RAVE sample did not contain many halo stars. The elongation
of the stream in the RAVE sample, however, could be a hint that actually more
than one stream on different orbital planes has been projected onto the Galactic
midplane.} The stars of this stream possess high vertical and low radial velocity
components, $\langle{W}\rangle=121\pm2$ km s$^{-1}$ and
$\langle{U}\rangle=24\pm2$ km s$^{-1}$ \citepalias[\S~6]{kle08}. Their
$V$-velocities range from $-180~ \text{km s}^{-1} \lesssim V \lesssim -140$ km
s$^{-1}$, so their orbital plane is inclined at an angle of
$\nu\approx30^\circ$. Indeed, in the $\nu$-range $15^\circ$-$45^\circ$ there
exist overdensities very close to the predicted position of the `RAVE' stream:
one in sub-sample s1, at $V_\text{az}\approx140$ km s$^{-1}$, which we labeled
as `R1' (Figure~\ref{fig:wl-s1}), and one in sub-samples s2 and s3, at lower
azimuthal velocities of $V_\text{az}\approx120$ km s$^{-1}$, which we labeled
`R2' (Figures~\ref{fig:wl-s2} and~\ref{fig:wl-s3}). The fact that `R2' is
located at the same position in s2 and s3 makes it unlikely that it arises
from Poisson noise. The possibility exists that `R1' and `R2', which possess
slightly different angular momenta, are causally connected, perhaps
resembling two distinct streams from a disrupted satellite.

To test this hypothesis we proceed as follows. We first examine the distribution
of $\nu$-angles for all stars that lie at the positions of `R1' and `R2' in
$(V_\text{az},V_{\Delta \text{E}})$-space, in order to check which orbital
inclinations can be assigned to the overdensities, and if they differ for `R1'
and `R2'. Then we can pick stream member stars according to their positions in
$V_\text{az},V_{\Delta \text{E}},$ and $\nu$, and look at their
metallicty distributions.

\begin{figure}[!ht]\centering
	\subfloat['R1']
		{\includegraphics[width=0.45\textwidth]{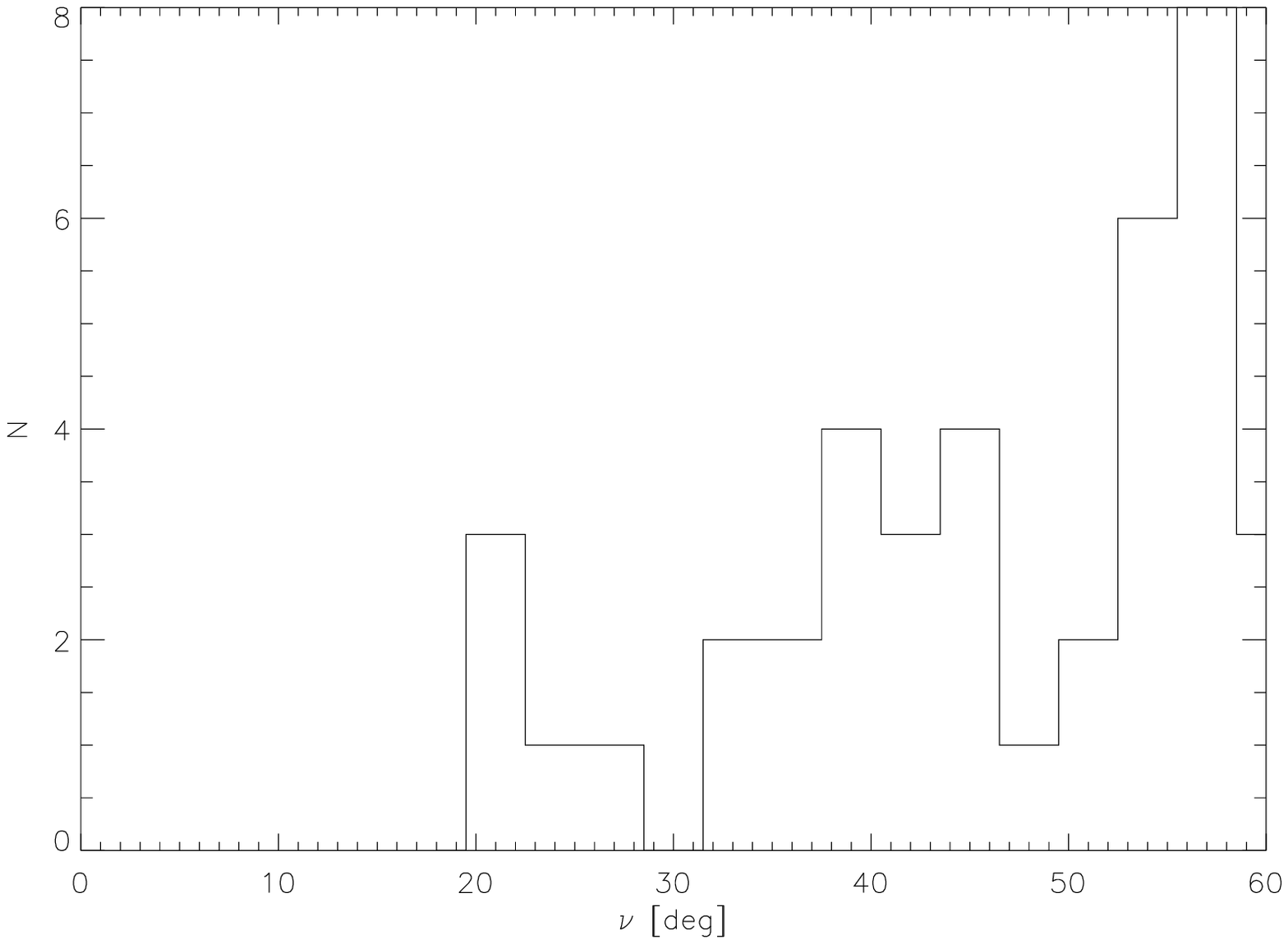}}
	\subfloat['R2']
	  {\includegraphics[width=0.45\textwidth]{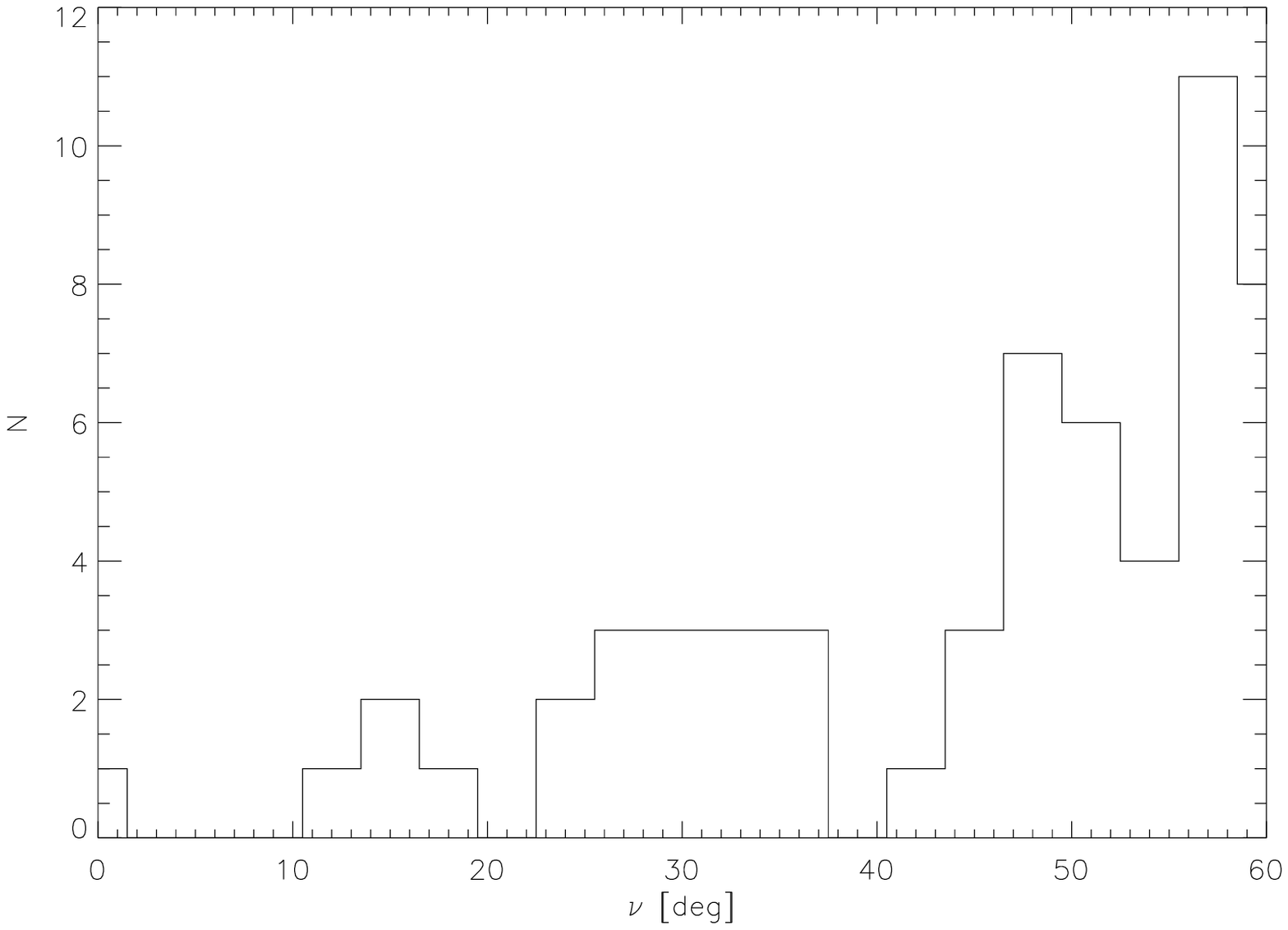}}
        \caption[`R1' and `R2': $\nu$ distributions]{Distribution of orbital
        inclinations $\nu$ of stars that lie at the same $(V_\text{az},V_{\Delta
        \text{E}})$-position as the red contoured peaks of the overdensities
        `R1' in Figure~\ref{fig:wl-s1} (a) and `R2' in Figure~\ref{fig:wl-s2}
        (b). The distribution for `R1' is not clearly peaked around a distinct
        orbital polar angle, while `R2' is centered at $\nu\approx30^\circ$. The
        $\nu$-range of 0-60$^\circ$ is part of the preliminary condition for
        putative member stars of `R1' and `R2'.}\label{fig:RAVEnus}
\end{figure}

We select the $(V_\text{az},V_{\Delta \text{E}})$-position of the two streams in
the following manner. For the stream at $V_\text{az}\approx140$ km s$^{-1}$, `R1',
we require $\lvert(V_\text{az},V_{\Delta \text{E}})-(140,120)\rvert\leq(30,30)$
km s$^{-1}$, $\lvert\nu-30^\circ\rvert\leq30^\circ$, and a value of the wavelet
transform in s1 of at least 90\% of its maximum value in that $\nu$-range (which
corresponds to the red-colored contours in
Figures~\ref{fig:wl-s1}-\ref{fig:wl-s4}). We also add stars from sub-samples s2
and s3 that lie in this region of $(V_\text{az},V_{\Delta \text{E}})$-space as
possible stream-members at lower metallicities, although their signature is not
visible in the significance maps. For the stream `R2', at $V_\text{az}\approx120$
km s$^{-1}$, we require $\lvert(V_\text{az},V_{\Delta \text{E}})-(120,150)
\rvert\leq(30,30)$ km s$^{-1}$, $\lvert\nu-30^\circ\rvert\leq30^\circ$, and a
value of the wavelet transform in s2 of at least 90\% of its maximum value in
that $\nu$-range. We also add stars from s1 that lie at this position as
possible stream members of higher metallicity. The distribution of $\nu$-angles
is shown in Figure~\ref{fig:RAVEnus}

\begin{figure}[!ht]\centering
	\subfloat[`R1']
		{\includegraphics[width=0.45\textwidth]{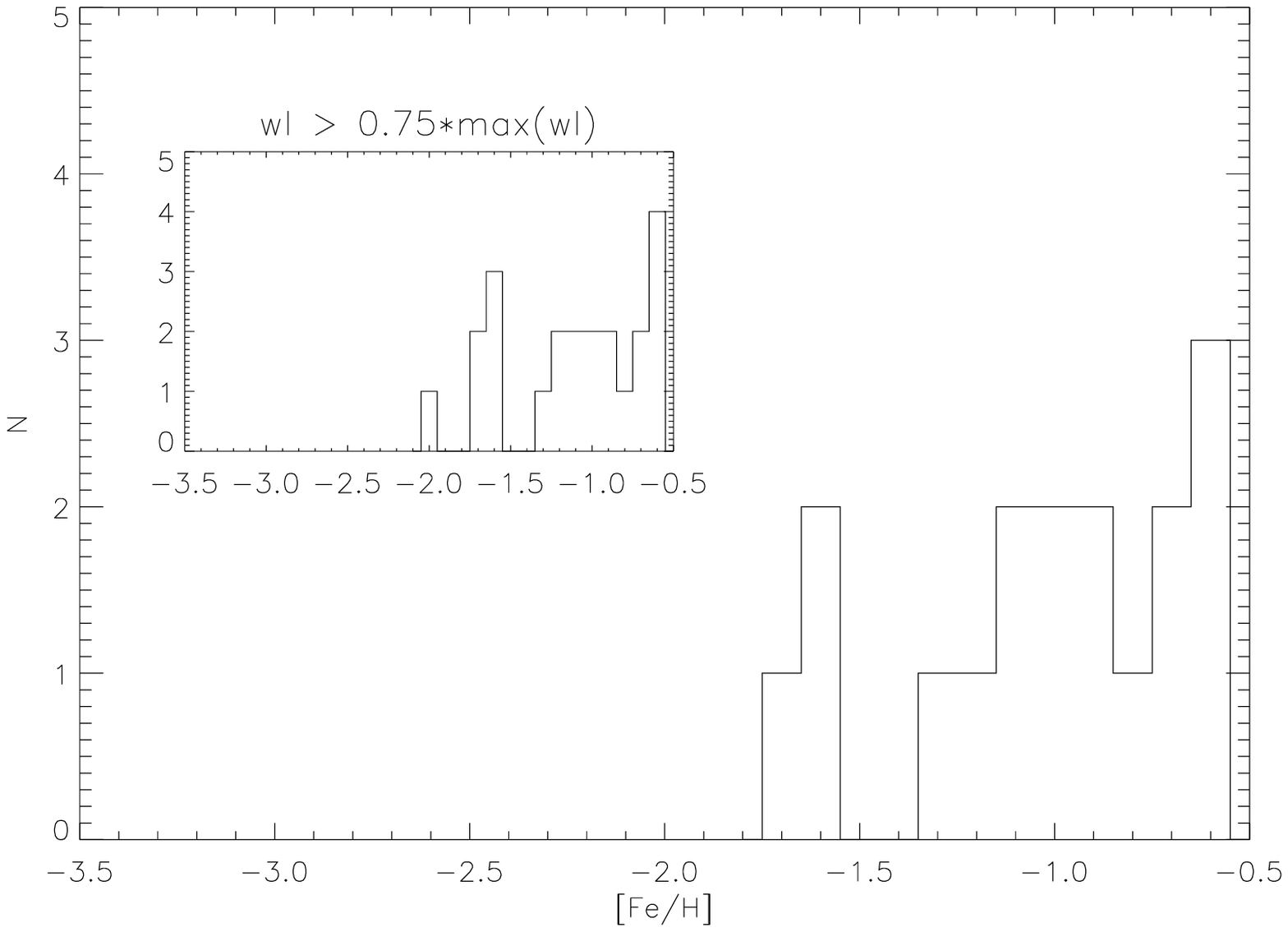}}
	\subfloat[`R2']
	  {\includegraphics[width=0.45\textwidth]{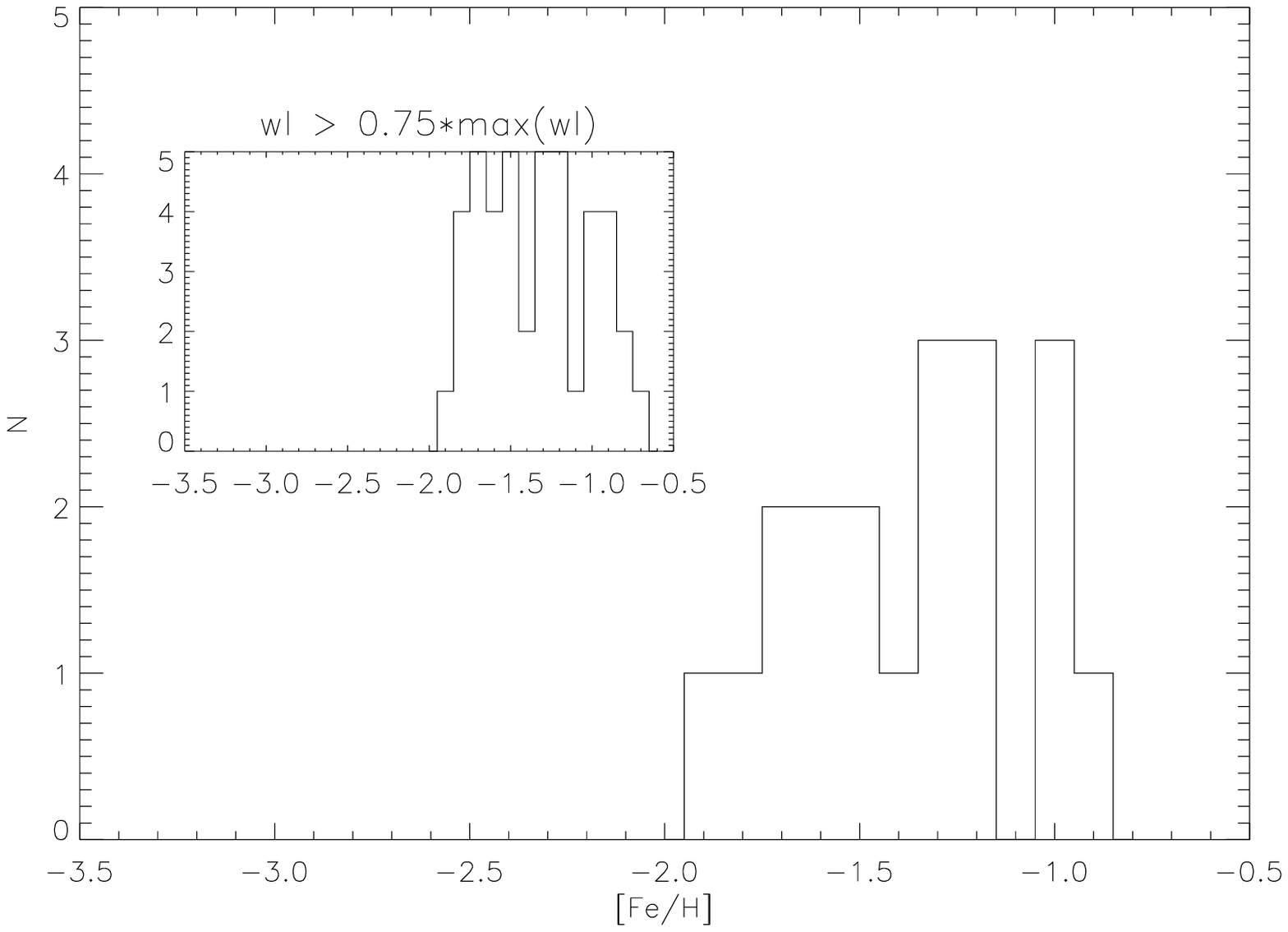}}
        \caption[`R1' and `R2': {[Fe/H]} distributions]{\textit{Large panels}:
        [Fe/H] distribution of stars that lie at the same $(V_\text{az},
        V_{\Delta \text{E}})$-position as the red contoured peaks
        (wl$=0.9\cdot$wl$_\text{max}$) of the overdensities `R1' in
        Figure~\ref{fig:wl-s1} (a) and `R2' in Figure~\ref{fig:wl-s2} (b). `R1'
        seems to peak at lower metallicity than `R2'. \textit{Small Panels}:
        Same as the large panels, but now stars have been selected from a larger
        region confined through wl$=0.75\cdot$wl$_\text{max}$. The difference
        between both [Fe/H] distributions is more
        pronounced.}\label{fig:RAVEfehs}
\end{figure}

From inspection of Figure~\ref{fig:RAVEnus} we cannot clearly assign an orbital
polar angle to the stream `R1'; stars distributed around $\nu\approx25^\circ$
and $\nu\approx40^\circ$ both contribute to the signal in the wavelet transform
in Figure~\ref{fig:wl-s1}. In contrast, at the position of `R2' there is a group
of stars that clump around $\nu=30^\circ$, which is the polar angle of
the `RAVE' stream. Assuming that `R1' and `R2' stem from the same progenitor, it
is possible that the difference in the $\nu$-distibutions is caused by
accelerations and decelerations of the stream stars by their precursor
object's potential \citep{choi07}. However, in this case they still should
exhibit the same metallicity distributions. The fact that `R1' is mostly present
in the sub-sample s1, while `R2' shows up in s2 and s3, suggests that the
[Fe/H] distributions are different. Figure~\ref{fig:RAVEfehs} compares
the [Fe/H] distributions, where we now confine the putative stream members into
the $\nu$-range $\lvert\nu-30^\circ\rvert\leq15^\circ$. In the large panels the
stars are further selected to lie at the position where the wavelet transform
takes on at least 90\% of its maximum value in this $\nu$-range, while we lessen
this reqirement to 75\% in the small panels to obtain more stars. 

The two [Fe/H] distributions are not compatible with the hypothesis that both
streams originate from the same precursor object. The [Fe/H] distribution of
`R2' peaks at metallicities between $-1.2$ and $-1.8$, and the stream does not
seem to contain stars more metal-poor than [Fe/H]$=-0.5$. On the other hand,
`R1' has one peak at [Fe/H]$\approx-1.6$, and a broad plateau possibly
continuing beyond [Fe/H]$=-0.5$. We have checked for a correlation
between the double-peaked $\nu$ distribution and the [Fe/H] distribution of
`R1', but it does not exist. We conclude from the
$\nu$- and [Fe/H] distributions that a tidal origin of `R1' is ruled out. `R1'
and `R2' are not correlated, in the sense that they originate from a single
progenitor, and the high significance of `R1' ($\sigma\approx8$) could be a
result of the small variance of our smooth reference model in this region of
$(V_\text{az},V_{\Delta\text{E}},\nu)$-space. 

\begin{figure}[!ht]\centering
	  \includegraphics[width=0.75\textwidth]{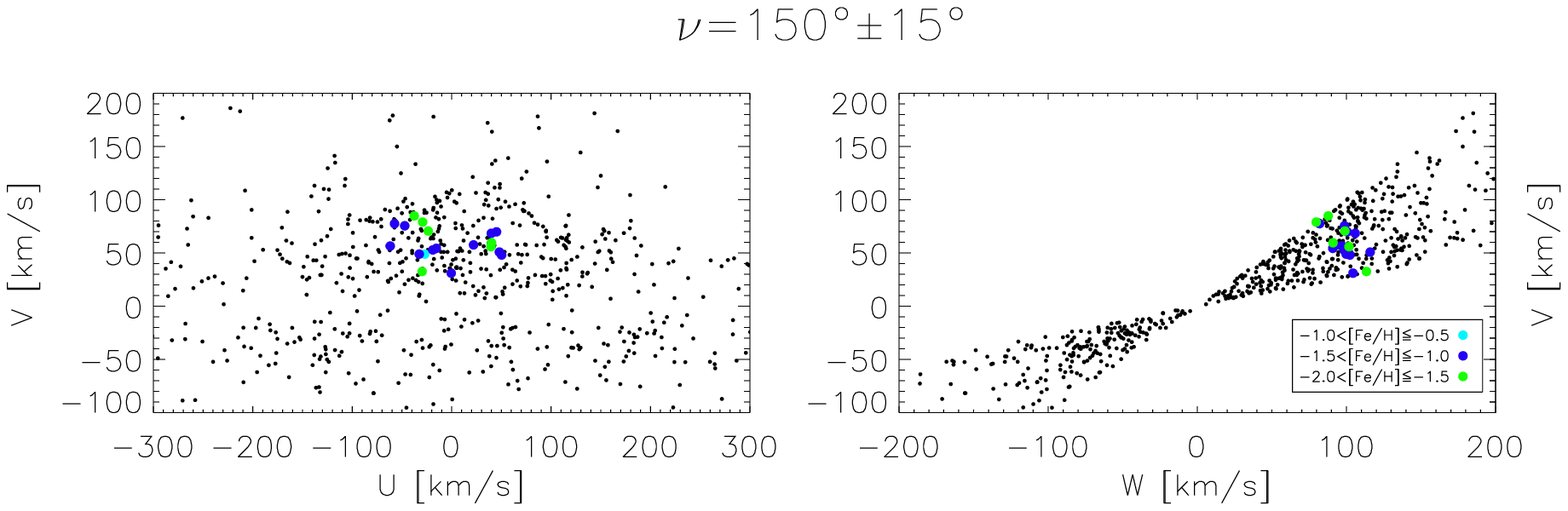}
        \caption['R2': ($U,V,W$) distribution]{Distribution of putative members
        of the stream discovered in RAVE data \citepalias{kle08} in ($U,V,W$).
        Note the banana shaped $(U,V)$ distribution centered at $U=0$,
        indicating that the stream stars are near their orbital
        apocenters.}\label{fig:f20}
\end{figure}

We retain `R2' as a likely tidal stream candidate and show its ($U,V,W$)
distribution in Figure~\ref{fig:f20}. Here, light blue dots represent stars in the
metallicity range $-1.0<\text{Fe/H}\leq-0.5$, blue dots for stars with
$-1.5<\text{Fe/H}\leq-1.0$, while green dots correspond to stars in the range
$-2.0<\text{Fe/H}\leq-1.5$. The small black dots are stars in the range
$-2.0<\text{Fe/H}\leq-0.5$ and $\lvert\nu-30^\circ\rvert\leq15^\circ$, and are
displayed as the background population. The stars approximately show a
``banana'' shaped distribution in $(U, V)$, which is typical for tidal streams near their
apocenters \citep{helm06}. The banana shape results from the condition
$V_{\Delta\text{E}}=\text{constant}$, which describes an ellipse in $(U,
V_\text{az})$, that is, in the radial and azimuthal velocity in the orbital
plane of the stream. If the distribution in $W$ is sufficiently narrow, this
shape also appears in the $(U,V)$ distribution.

The $U$,$V$, and $W$ velocities are consistent with those found in the RAVE sample: $(\langle U\rangle,\langle V\rangle,\langle
W\rangle)=(-5\pm13,59\pm5,98\pm3)$ km s$^{-1}$. We have good reason to believe
that we have rediscovered their proposed new stream. The
metallicity distribution, shown in Figure~\ref{fig:RAVEfehs}(b), suggests that
the stream consists of stars mainly in the range
$-1.8\lesssim\text{[Fe/H]}\lesssim-1.2$. 

\subsection{A New Stream Candidate}

\begin{figure}[!ht]\centering
	\includegraphics[width=0.8\textwidth]{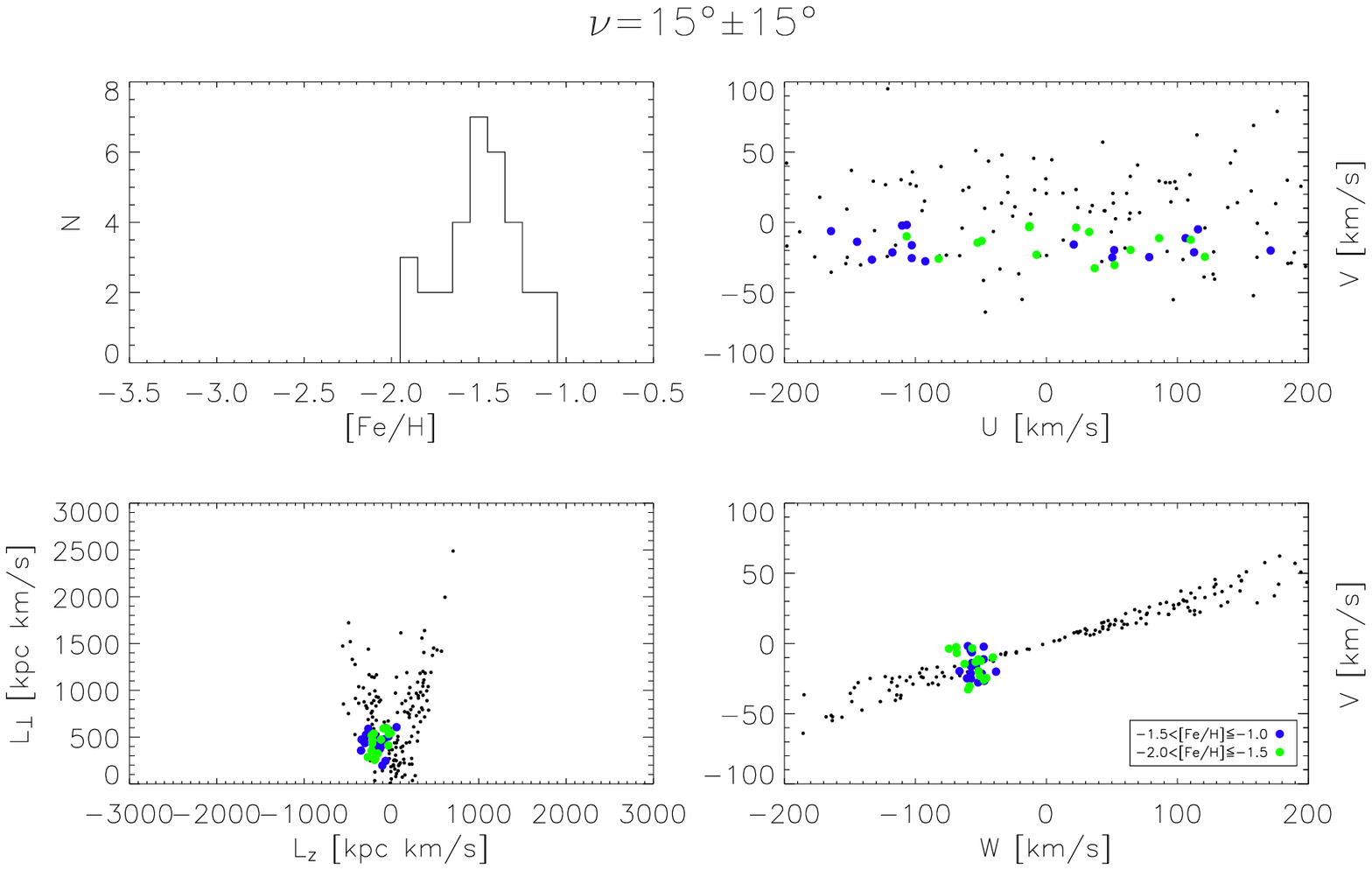}
        \caption[`C1': {[Fe/H]}-, ($L_z,L_\perp$)- and ($U,V,W$) distributions]
        {Distribution of members of the stream candidate `C1' in metallicity
        [Fe/H], angular momentum ($L_z,L_\perp$), and ($U,V,W$). Blue dots show
        stars in the range $-1.5<\text{Fe/H}\leq-1.0$, while green dots stars
        represent stars with $-2.0<\text{Fe/H}\leq-1.5$. The small black dots
        are all stars in our sample with $-2.0<\text{Fe/H}\leq-1.0$ and
        $\lvert\nu-15^\circ\rvert\leq15^\circ$. The $(U,V)$ distribution is
        symmetric around $U=0$, indicating that the stream stars are well-mixed
        and near their orbital apocenters.}\label{fig:C1stream}
\end{figure}

Centered at even lower orbital polar angles than the `RAVE' stream in sub-sample
s2 ($-1.5<\text{[Fe/H]}<-1.0$) (Figure~\ref{fig:wl-s2}), we find an overdensity
of stars around $(V_\text{az},V_{\Delta \text{E}})\approx(-60,410)$. This
overdensity extends towards the sub-sample s3, so it is unlikely to be caused by
Poisson noise. Because we do not know of any stream in the literature with such
kinematics, we have labeled this overdensity `C1' for our first new stream
candidate. The significance of this feature is $\sigma\approx2.9$. We analyze
its [Fe/H], angular momentum, and velocity distribution in
Figure~\ref{fig:C1stream}.

Because the velocity distribution is symmetric around $U=0$, these stars must be
moving towards and away from their apocenters. The typical banana shape is
indicated in the ($U,V$) distribution, but not perfectly so. Because of the high
polar angle of the orbit, the banana shape in $(U,V_\text{az})$ (which is predicted
by the condition $V_{\Delta \text{E}}=$constant) does not perfectly translate into
$(U,V)$. The metallicity distribution is roughly symmetric around [Fe/H]$=-1.5$
and hints towards the distribution of tidal debris from a single metal-poor
progenitor. We propose C1 to be a newly discovered halo stream passing through
the solar vicinity.

\subsection{Two Related Streams?} 
\label{sec:S3C2}
The retrograde stream labeled `S$_3$' was discovered by \citet{det07} as an
overdensity centered at $(V_\text{az},V_{\Delta \text{E}},\nu)=(-100 \text{km
s}^{-1},470 \text{km s}^{-1},170^\circ)$. We also find an overdensity of stars
at this position in sub-sample s2 ($-1.5<\text{[Fe/H]}<-1.0$), but peaked at
$\nu\approx155^\circ$. Nevertheless, we identify this overdensity with `S$_3$',
because from Figure~3 in \citet{det07} we can see that the wavelet contours of
their feature seem to extend towards $\nu\approx155^\circ$. 

\begin{figure}[!ht]\centering
	\includegraphics[width=0.75\textwidth]{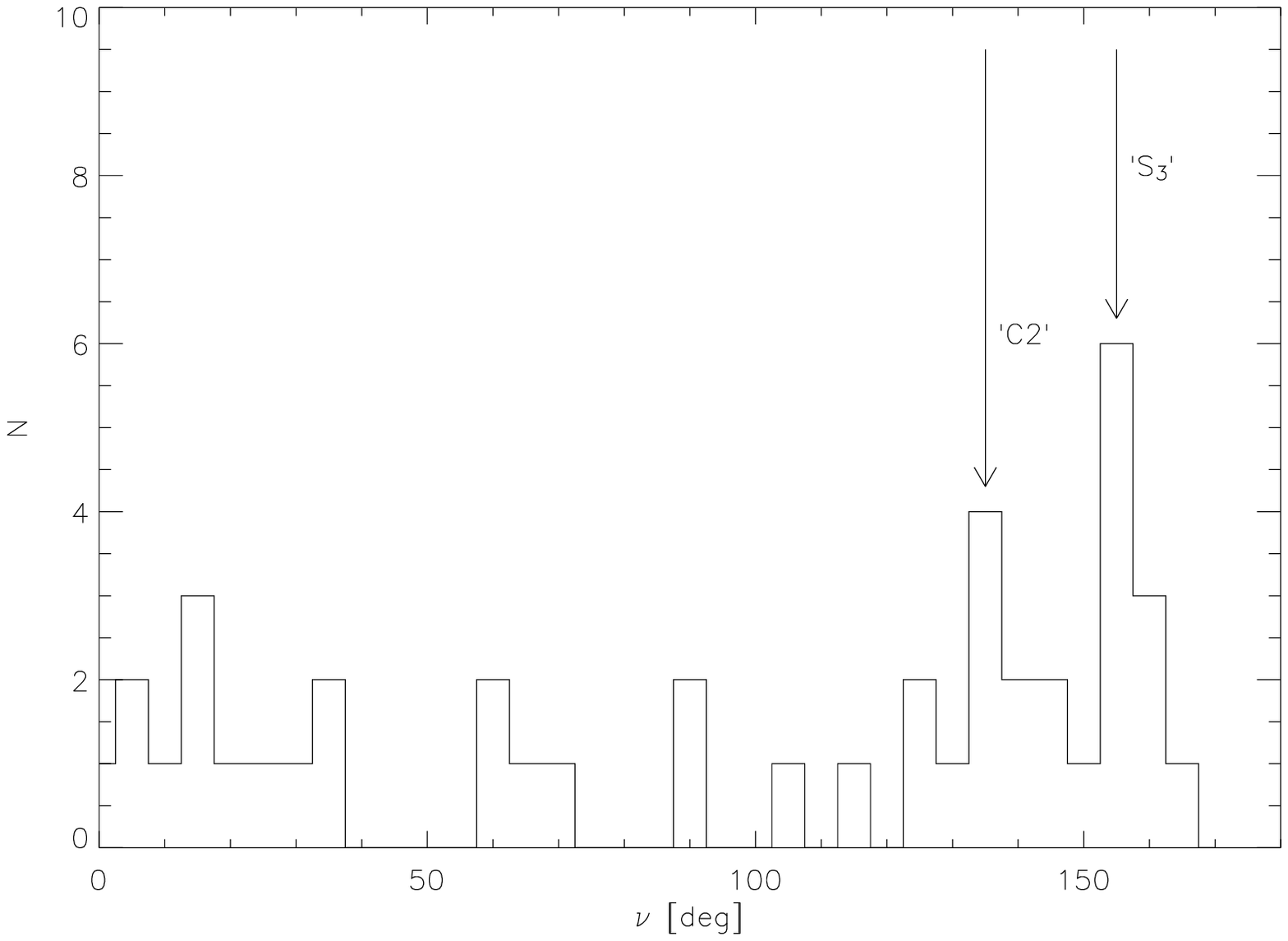}
        \caption[$\nu$ distribution of stars in s2 with the same $(V_\text{az},
        V_{\Delta\text{E}})$ as the overdensity `S3']{Distribution of orbital
        inclinations $\nu$ for all stars in sub-sample s2 that occupy the same
        region in $(V_\text{az},V_{\Delta\text{E}})$-space as the overdensity
        labeled `S$_3$' in Figure~\ref{fig:wl-s2}, $\nu$-slice
        $135^\circ$-$165^\circ$. The location of the overdensity is selected
        from the appropriate ranges in $V_\text{az}$ and $V_{\Delta\text{E}}$
        and the condition that the wavelet transform has at least 50\% of its
        maximum value (to select the green-framed ``bump''.)}
\label{fig:S3nuhist}
\end{figure}

In the $\nu$-slice $120^\circ$-$150^\circ$ we detect another overdensity at
nearly the same $(V_\text{az},V_{\Delta \text{E}})$ values as `S$_3$', to
which we assign the name `C2'. This corresponds to a second peak in the $\nu$ distribution of stars in the $(V_\text{az},V_{\Delta \text{E}})$ range centered
at $\approx(-105,480)$ km s$^{-1}$. We suspect that both
`C2' and `S$_3$' contribute to the high signal of the wavelet transform in the
$\nu$-slice $135^\circ$-$165^\circ$ in Figure~\ref{fig:wl-s2}. This is confirmed in
Figure~\ref{fig:S3nuhist}, where we have plotted the $\nu$ distribution of all
stars that are located at the same $(V_\text{az},V_{\Delta \text{E}})$ position
as the overdense region in this $\nu$-slice. The two peaks at
$\nu\approx155^\circ$ and $\nu\approx135^\circ$ correspond to the streams
`S$_3$' and `C2', respectively.

Figure~\ref{fig:S3nuhist} seems to suggest that `S$_3$' and `C2' are two distinct streams that move with nearly the
same orbital inclinations and eccentricities, but on different orbital planes
that differ by only $\sim20^\circ$. Such a double-peaked $\nu$ distribution could
hint towards two tidal streams lost at different times from a progenitor whose
orbital plane has precessed slightly during many orbits in the Milky Way. However, on a more careful look at Figure~\ref{fig:S3nuhist}
we see that only a handful of stars are in each distinct peak. The apparent bimodality already becomes much less prominent when we increase
the binsize of the histogram from 5$^\circ$ to 7$^\circ$. If we consider all sources of errors and noise ([Fe/H]-determination,
possible systematic distance errors, proper motion errors, unresolved binaries,
Poisson noise), could it be that we're looking at a single stream that has
just been smeared out in $\nu$ space by the errors?

We test this hypotheses by analyzing the kinematical and chemical properties of
both features.

\begin{figure}[!ht]\centering
		\includegraphics[width=0.8\textwidth]{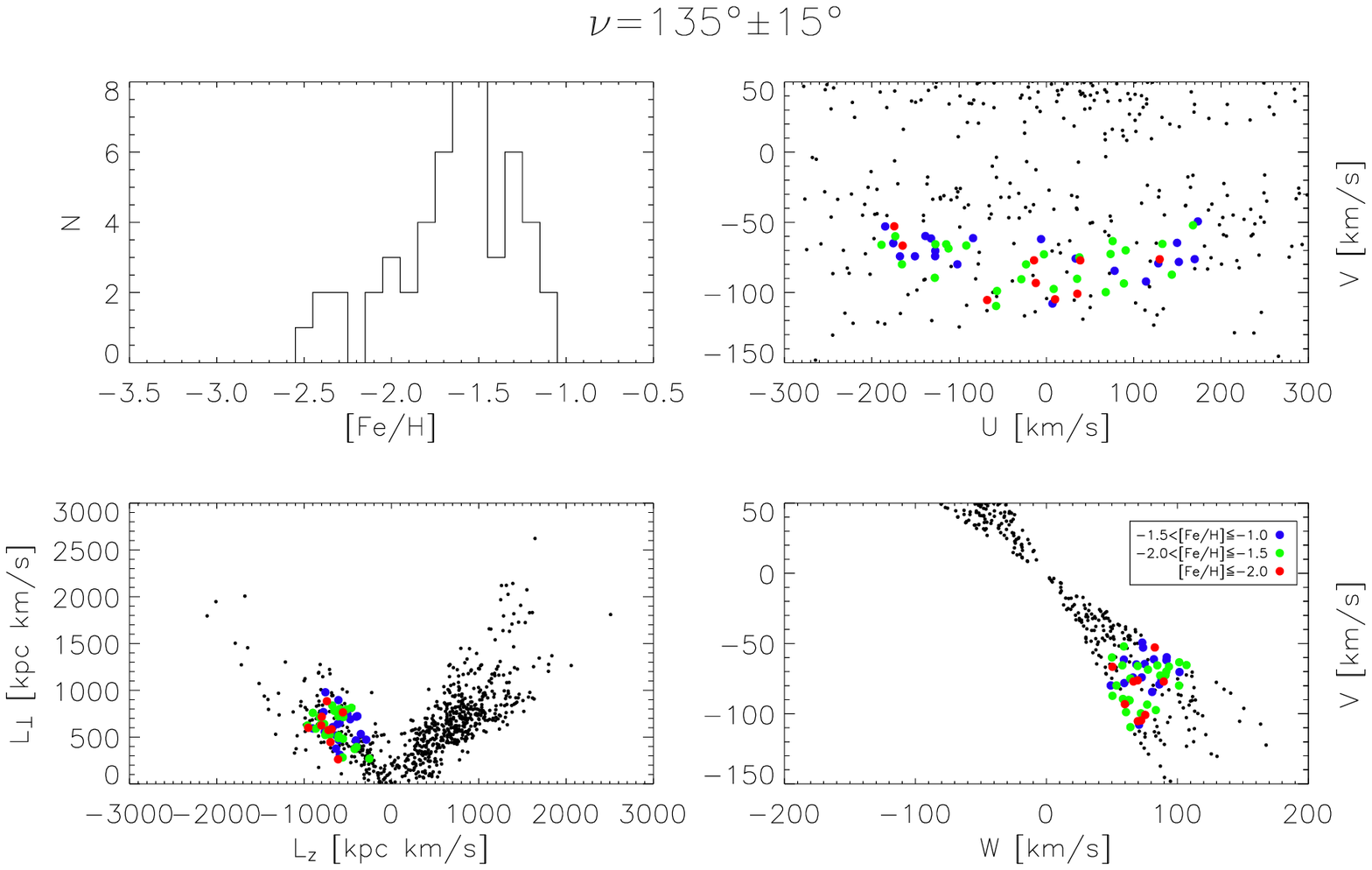}
        \caption['C2': {[Fe/H]}, ($L_z,L_\perp$) and ($U,V,W$) distributions]
        {[Fe/H], ($L_z,L_\perp$), and ($U,V,W$) distributions for stars belonging
        to the stream candidate `C2'. Light blue dots show stars in the
        metallicity range
        $-1.0<\text{Fe/H}\leq-0.5$, dark blue dots in the range
        $-1.5<\text{Fe/H}\leq-1.0$, and green dots stars with
        $-2.0<\text{Fe/H}\leq-1.5$. The small black dots are all stars in our
        sample that occupy the same [Fe/H]- and $\nu$-ranges as the member stars
        of `C2'.}\label{fig:f23}
\end{figure}

\begin{figure}[!ht]\centering
		\includegraphics[width=0.8\textwidth]{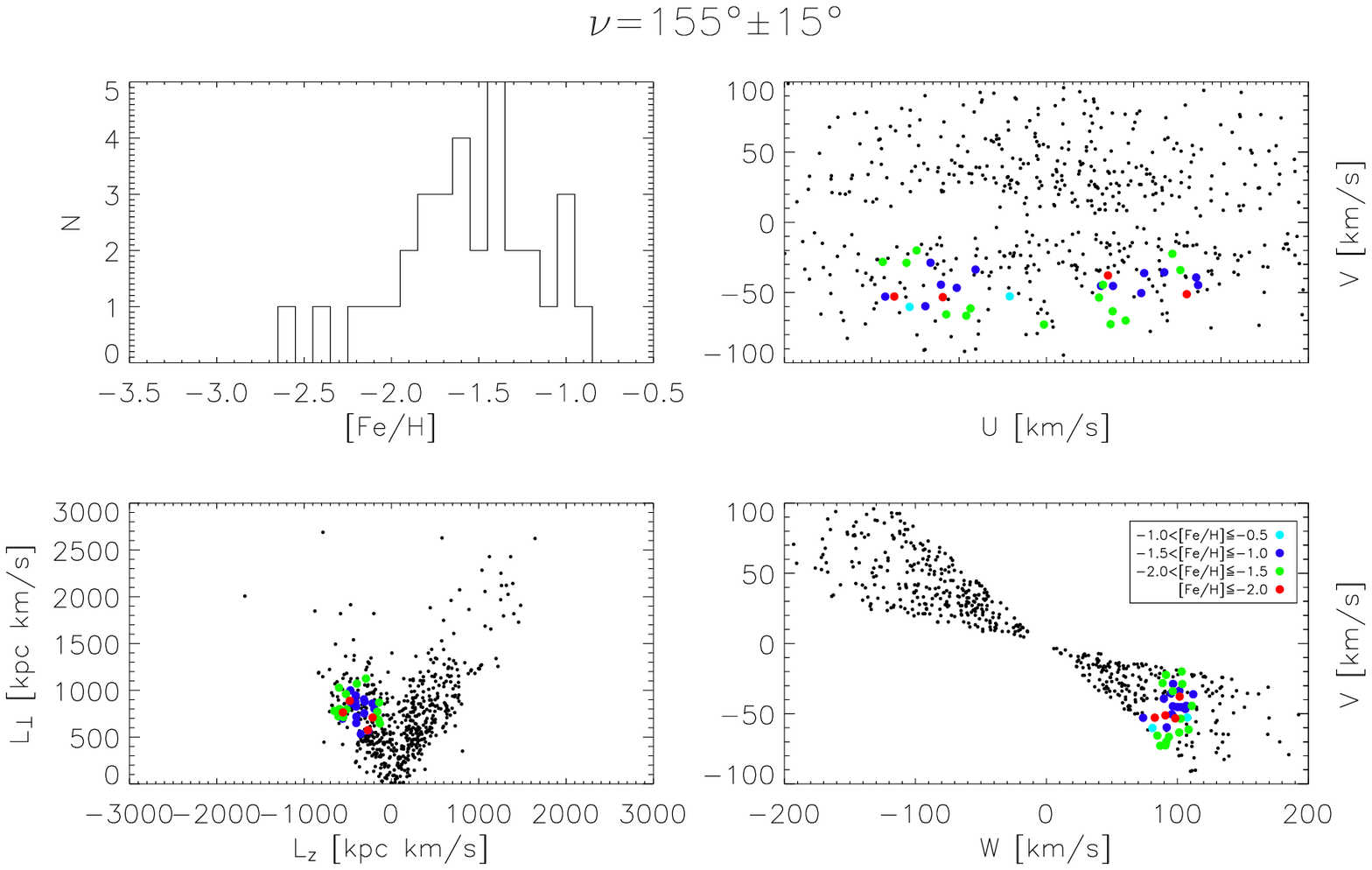}
        \caption[`S$_3$': {[Fe/H]}, ($L_z,L_\perp$) and ($U,V,W$)
        distributions]{[Fe/H], ($L_z,L_\perp$), and ($U,V,W$) distributions for
        stars belonging to the stream `S$_3$'. The colors have the same meaning
        as in Figure~\ref{fig:f23}.}\label{fig:f24}
\end{figure}

Figures~\ref{fig:f23} and~\ref{fig:f24} show the [Fe/H], $(L_z,L_\perp)$
and $(U,V,W)$ distributions for `C2' and `S$_3$', respectively. Although it
seems that `C2' does not contain stars in the range $-2.0<\text{[Fe/H]}\leq-1.5$
from the ``gap" in the wavelet transform contours in Figure~\ref{fig:wl-s3}, we
include stars from sub-samples s3 and s4 that lie in the same region as the
putative stream members from s2. This is justified from the [Fe/H] distribution
shown in Figure~\ref{fig:f23}, which peaks at [Fe/H] $\approx-1.5$, and falls off along
a tail towards lower metallicities. It resembles the distribution of a coeval
tidally disrupted stellar population. We do the same for `S3', and include all
stars in the range $\text{[Fe/H]}\leq-0.5$ and
$\lvert\nu-155^\circ\rvert<15^\circ$. This stream also peaks around
[Fe/H]$\approx-1.5$ (Figure~\ref{fig:f24}). We note that through these
selection criteria nine stars appear in both `C2' and `S$_3$'. 


The banana-shaped ($U,V$) distribution for the stream `C2' indicates that its
stars are near their orbital apocenters. The velocity distribution of `S$_3$' is only
slightly different. The stars appear to be not as close to their apocenters as the
`C2' stars. Both the $(U,V,W)$ and [Fe/H] as well as the similar $L_z$ distributions seem to support the
hypothesis that `C2' and `S$_3$' are in fact one single stellar stream. Taking into account
the typical statistical velocity errors discussed in \S~\ref{sec:ComparisonWithFiducials}, we find that
stars at such velocities as those of `C2' could well be misplaced by $\pm5^\circ$ in $\nu$ and $\pm5$ km s$^{-1}$ in $V_\text{az}$ and $V_{\Delta\text{E}}$, which
would lead to a smearing out of one single stream in $\nu$ space \citep[Fig.~5.8(b)][see also]{kle08a}.

We conclude that there exists evidence from the $(U,V,W)$ and [Fe/H] distributions that `C2' and `S$_3$' belong to one single stream which might has been smeared out in $\nu$ space and -- to a lesser degree -- in $(V_\text{az},V_{\Delta \text{E}})$ space. In addition, the $L_z$ distributions are similar enough to support this hypotheses. 

\clearpage

\subsection{The Helmi Stream}

We again examine the metallicity range of sub-sample s2,
$-1.5<\text{Fe/H}\leq-1.0$, and find an overdensity of stars on a highly
prograde orbit inclined at $\nu\approx150^\circ$. This is the stream discovered
by \citet{helm99}, located at $(V_\text{az},V_{\Delta \text{E}},\nu)
=(300,120,150^\circ)$, in very good agreement with the signal of this stream in
the dataset of \citet{det07} (see Table~\ref{tab:stcoords}). The stream, labeled
`H99', was originally discovered as an overdensity of stars in angular momentum
space. Later, \citet{chi00} confirmed the existence of this stream in their own
dataset and identified a possible extension towards higher azimuthal rotation.
Although their dataset was approximately three times the size of the sample used
by \citet{helm99}, the number of stream stars ($N=10$) stayed constant. They
hypothesized that the `H99' stream could be related to the ``trail" extension, if
this ``trail'' gained angular momentum from the interaction of the progenitor
with the Milky Way's gravitational potential. 

\begin{figure}[!ht]\centering
	\includegraphics[width=0.75\textwidth]{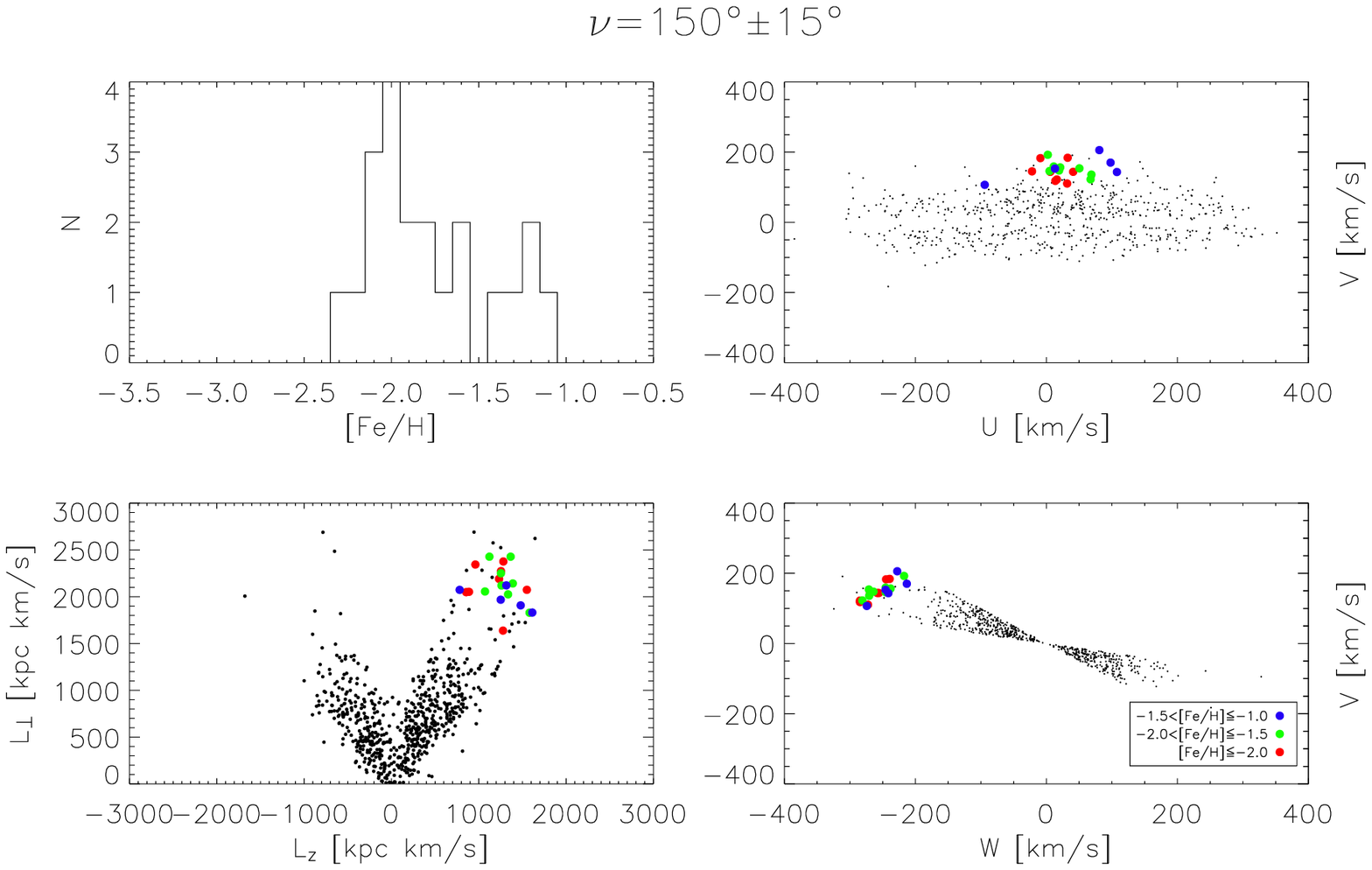}
        \caption['H99': {[Fe/H]}-, ($L_z,L_\perp$)- and ($U,V,W$)
        distributions]{Distributions of members of the `H99' stream in [Fe/H],
        ($L_z,L_\perp$), and ($U,V,W$). Blue dots show stars in the metallicity range
        $-1.5<\text{Fe/H}\leq-1.0$, green dots stars with
        $-2.0<\text{Fe/H}\leq-1.5$, and red dots stars with
        $\text{Fe/H}\leq-2.0$. The small black dots are all stars in our sample
        with $\text{Fe/H}\leq-1.0$ and $\lvert\nu-150^\circ\rvert\leq15^\circ$.
        The $(U,V,W)$ distribution is in very good agreement to that shown by
        \citet[][their Figure 2]{helm99}.}\label{fig:f27}
\end{figure}

Figure~\ref{fig:f27} shows the [Fe/H]-, ($L_z,L_\perp$)- and ($U,V,W$) distribution of all stars
that we assign to the `H99' stream. The stream extends
towards lower metallicities and is highly significant in the sub-sample s4. The
$(U,V,W)$- and $(L_z,L_\perp)$ distributions agree very well with those in the
original work of \citet[][their Figure 2]{helm99}. The [Fe/H] distribution peaks
at [Fe/H] $\approx-2.0$ and does not extend beyond [Fe/H] $\approx-2.3$. The range
of [Fe/H]-values that we find for the `H99' stream agrees well with the
[Fe/H]-values of the `H99' stream members given by \citet{kep07}, although they
report two stream members with [Fe/H] $< -2.3$. The number of stars in our sample
that we identify as `H99' members is $N=21$, approximately double that of
previous studies.  However, the number is too small to account for as much as
one-tenth of the halo stars that are currently present in the solar
neighborhood, as suggested by \citet{helm99}.  

\subsection{More Substructure at Very Low Metallicities}
\label{sec:C3C4}

In the sub-samples s3 ($-2.0<\text{[Fe/H]}<-1.5$) and s4 ($\text{[Fe/H]}<-2.0$),
the amount of substructure increases further. We expect a contribution of only
$\sim10$\% thick-disk stars in s3 and 100\% halo stars in s4 \citep{chi00}.
While about 46\% of the stars in sub-sample s2 move on disk-like orbits
($75^\circ\leq\nu<105^\circ$), this number drops to 23\% for both s3 and s4. It
seems that also for stars more metal-poor than [Fe/H]$=-2$ a small fraction of
stars with (metal-weak) thick-disk-like kinematics remains constant.  

\begin{figure}[!ht]\centering
	\includegraphics[width=0.8\textwidth]{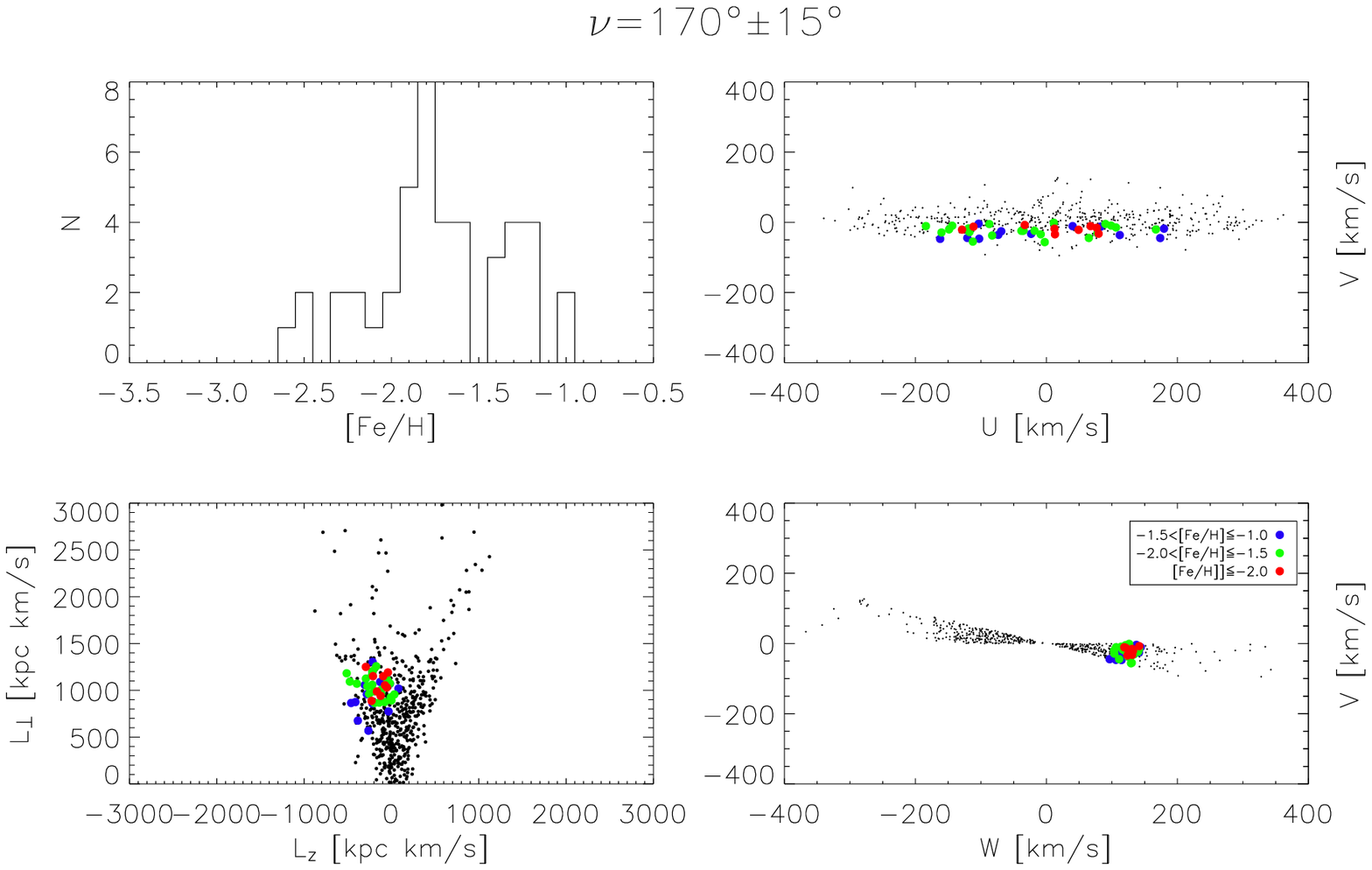}
        \caption['C3': {[Fe/H]}-, ($L_z,L_\perp$)- and ($U,V,W$)
        distributions]{Distribution of members of the `C3' stream candidate in
        [Fe/H], ($L_z,L_\perp$), and ($U,V,W$). Blue dots show stars in the
        metallicity range
        $-1.5<\text{Fe/H}\leq-1.0$, green dots stars with
        $-2.0<\text{Fe/H}\leq-1.5$, and red dots stars with
        $\text{Fe/H}\leq-2.0$. The small black dots are all stars in our sample
        with $\text{Fe/H}\leq-1.0$ and
        $\lvert\nu-170^\circ\rvert\leq15^\circ$.}\label{fig:C3stream}
\end{figure}

Besides the already discussed streams that are also visible in s3, `RAVE', `H99'
and `C1', we find an overdensity centered at $(V_\text{az},V_{\Delta \text{E}},
\nu)=(-130 \text{km s}^{-1},510 \text{km s}^{-1},170^\circ)$, which we label as
`C3'. It is located near the stream `S$_3$' and -- like `C2' -- may be related to
it. The peak in the $\nu$ distribution of stars in this region of $(V_\text{az},
V_{\Delta \text{E}})$-space at $\nu=170$ would fit exactly to the $\nu$-value
originally assigned to `S$_3$' by \citet{det07}. The [Fe/H]
distribution peaks at [Fe/H] $\approx -1.8$, and is consistent with that
of `S$_3$' or `C2' (Figure~\ref{fig:C3stream}). The velocity and $L_z$ distribution 
of `C3' seem to differ slightly from those of the `S$_3$/C2' stream. However, like in the case of `C2' and `S$_3$', 
if we consider all possible error sources the possibility exists that one single stream manifests itself in several adjacent features. 
In particular, we refer to \S~\ref{sec:systdisterr}, where we have shown that a systematic distance error of 10\% alone is able to
cause such an effect through changing the relative ``heights" of overdensities. In addition, as already noted in \S~\ref{sec6}, the confidence levels for
sub-structures the way we compute them depend on the priors we choose
for the Monte Carlo simulations (smooth halo). But given the case that the `S$_3$/C2' stream is there, its existence increases the number
of stars in that region of phase-space, which increases the level of
Poissonian fluctuations (as they are proportional to $\sqrt{N}$). When we divide
the residuals between a smooth background and these fluctuations by the expected sigma for the smooth background, it may lead to the
appearance of apparently highly significant multiple, adjacent, peaks
(while in fact they are a part of the same structure).

We gain confidence that `C3' is a not a feature created from Poisson noise from
the fact that the wavelet transform at `C3's position in $(V_\text{az},V_{\Delta
\text{E}})$-space has values greater than 90\% of its maximum in both
sub-samples s3 and s4 (Figures\ref{fig:wl-s3} and~\ref{fig:wl-s4}). Furthermore,
`C3' appears in the significance map of sub-sample s3 (Figure~\ref{fig:sig-s3})
at a significance level greater than 2.

\begin{figure}[!ht]\centering
\includegraphics[width=0.8\textwidth]{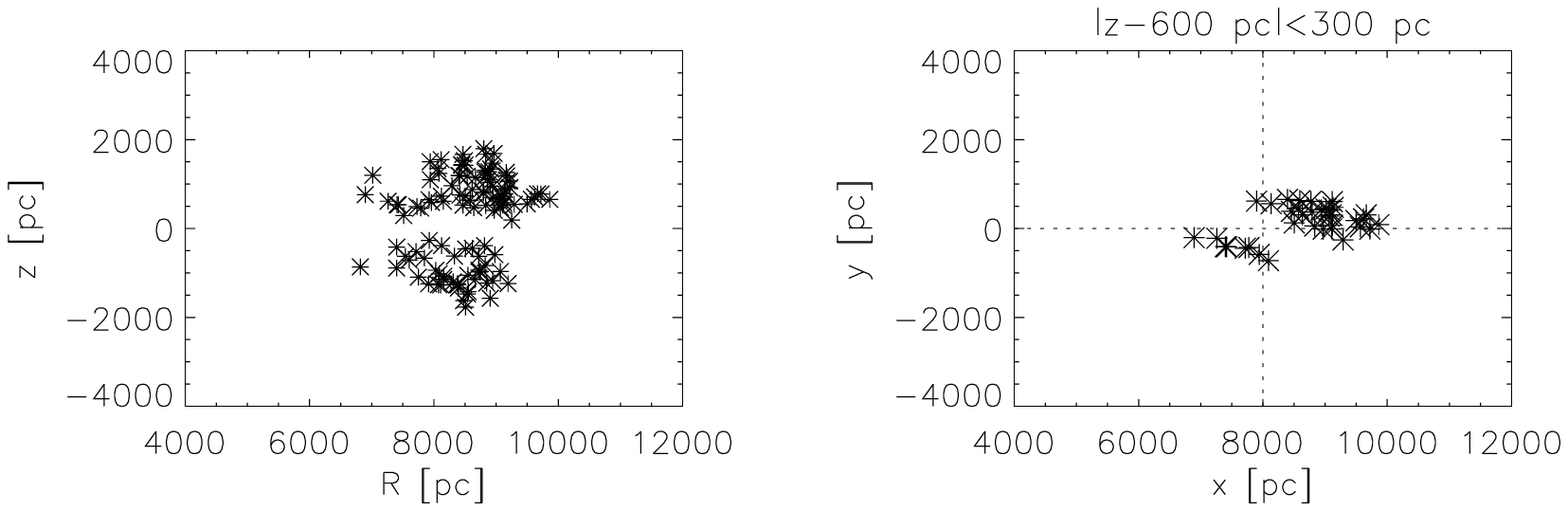}
        \caption[Spatial distribution of the `S$_3$/C2/C3' stream.]{Spatial distribution of stars which we identify as members of the composite stream 
        `S$_3$/C2/C3'. In the left panel we show all 97 stars in the $(R-z)$ plane, while in the right panel we concentrate on 
        a slice at $z=600$ pc, analog to Figure~27 in \citet{jur08}. Note the very similar location of the main overdensity in the $z=600$ pc slice.}\label{fig:f29}
\end{figure}

Given its proximity to the `S$_3$/C2' stream and the uncertainties in
our variables (in particular [Fe/H], systematic distance errors and Poisson noise),
we think that `C3' is part of that same stream. Taken together, the `S$_3$/C2/C3' stream consists of 97 stars
whose spatial distribution shows similarities with the thick-disk overdensity found by \citet{jur08} at $(R,z)\approx(9.5,0.8)$ kpc.
We show this in Figure~\ref{fig:f29}, where we have adopted the coordinate system used by \citeauthor{jur08} (with the x-axis pointing towards the Sun). A comparison to their Figure~27 (\textit{left and right panel}) reveals an intriguingly similar distribution in the Northern hemisphere plus a counterpart in the Southern hemisphere which -- although less prominent -- is also mapped by \citeauthor{jur08} in their Figure~26 (\textit{second row, right panel}). The $(x,y)$-distribution in the $z=600$ pc slice reveals a main clump, whose position matches the rectangular region in the right panel of \citeauthor{jur08}'s Figure~27. This is to our knowledge the first time that a 6D phase-space overdensity has been firstly identified by its kinematics. If the similarity with the \citeauthor{jur08} overdensity is not a coincidence, this would confirm this finding by an independent method. Further, the kinematics of the feature would rule out a ringlike feature of (thick) disk stars, instead supporting either a ``localized clumpy overdensity" (\citeauthor{jur08}) or a spatially coherent stellar stream passing the plane in the northern direction. However, our survey volume is not large enough to distinguish between these two possibilities.

We note that the `S$_3$/C2/C3' stream is the only stream for which we have found signs of a spatial coherence.   

\begin{figure}[!ht]\centering
\includegraphics[width=0.8\textwidth]{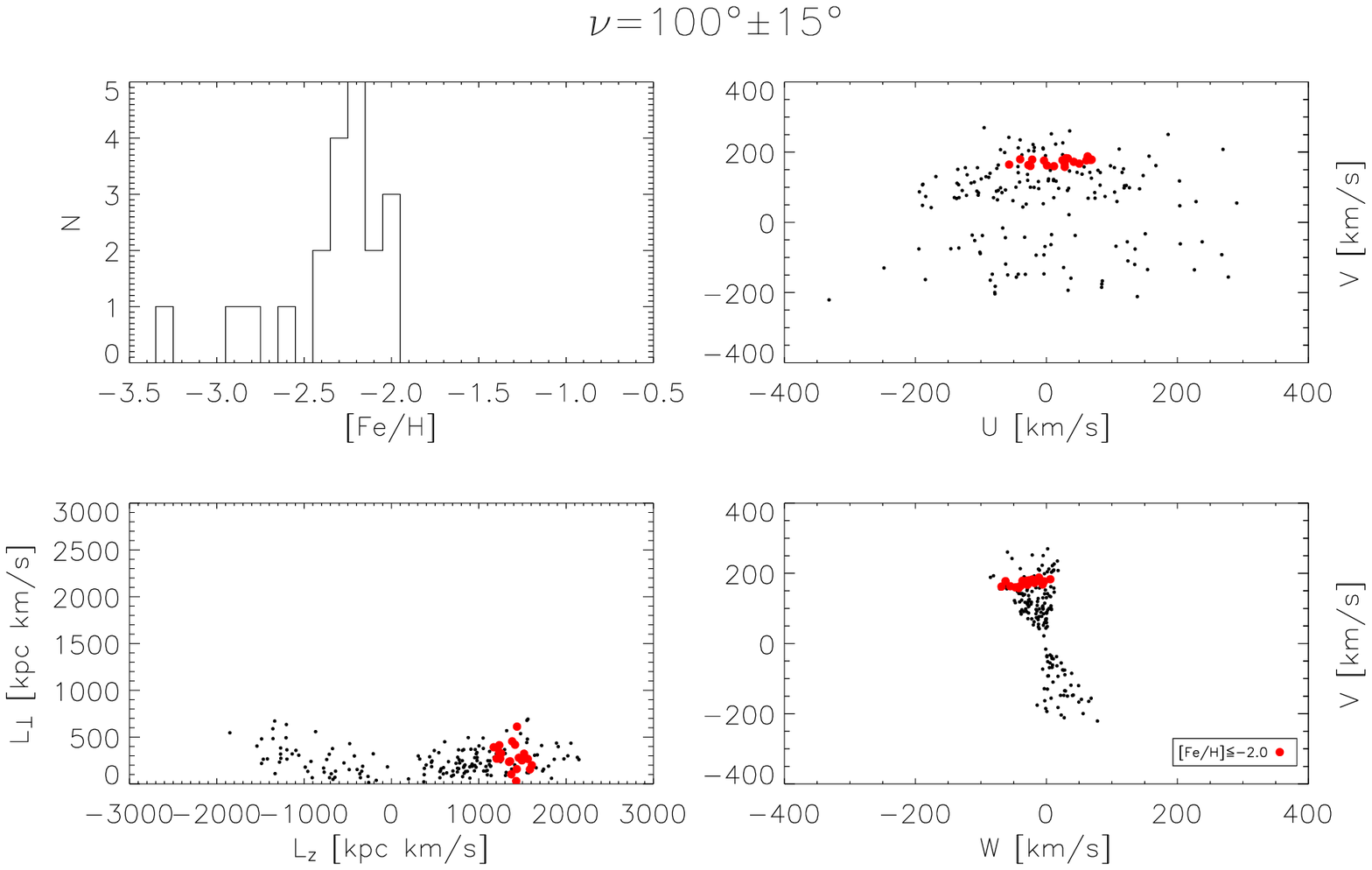}
        \caption['C4': {[Fe/H]}-, ($L_z,L_\perp$)- and ($U,V,W$)
        distributions]{Distribution of members of the `C4' stream candidate in
        metallicity [Fe/H], ($L_z,L_\perp$), and ($U,V,W$). The small black dots
        are all stars in our sample with $\text{Fe/H}\leq-2.0$ and
        $\lvert\nu-100^\circ\rvert\leq10^\circ$.}\label{fig:C4stream}
\end{figure}

\clearpage
In the most metal-poor sub-sample s4 ($\text{[Fe/H]}<-2.0$), where the fraction
of halo stars is close to 100\%, there exists a substantial amount of
substructure; it is difficult to identify any smooth component. However,
many of the overdensities in Figure~\ref{fig:wl-s4} consist of only a few stars,
and their reality is thus in doubt. We concentrate on a highly
significant ($\sigma=4.8$) feature located at $(V_\text{az},V_{\Delta \text{E}},
\nu) =(175,75,100^\circ)$ and labeled with `C4'. In this region of
$(V_\text{az}, V_{\Delta \text{E}},\nu,
\text[Fe/H])$-space a density enhancement of stars is not expected. The `C4'
candidate stream is
clumped around $(\nu,\text[Fe/H])\approx(100^\circ,-2.2)$, which clearly
distinguishes it from a smooth feature. Figure~\ref{fig:C4stream} shows
the metallicity, angular momentum- and ($U,V,W$) distribution of `C4'. 

The very low excentricity, $e=V_{\Delta
\text{E}}/\sqrt{2}V_\text{LSR}\simeq0.2$, of `C4' suggests that it belongs to
the metal-weak thick disk. However, according to \citet{chi00} even the
metal-weak tail of the thick disk should not contain stars as metal poor as
[Fe/H] $\lesssim-2.2$, where the [Fe/H] distribution of `C4' peaks . These
authors argued that the fraction of low-excentricity stars with [Fe/H]$\leq-2.2$
remains the same regardless of their height $\lvert z\rvert$, implying that they
purely belong to the halo. In addition, stars in the clump `C4' are not
distributed symmetrically around $\nu=90^\circ$, but are centered at an orbital
polar angle of roughly 100$^\circ$, with a longer tail towards higher
inclinations. 

On the other hand, the formation of the stellar halo and thick disk might
underlie a common cause -- the accretion of satellite galaxies. It has been
shown that the metal-weak tail of the thick disk could consist of tidal debris
stemming from a progenitor on a planar orbit that has been circularized prior to
disruption through dynamical friction \citep{quinn86,quinn93,abadi03b}. The very
low abundances of the `C4' stars favor this scenario. Even if these stars do
not belong to a single progenitor exclusively, we have found evidence that tidal
debris exists on disk-like orbits, giving further support for the hierarchical
build-up of the Milky Way and making `C4' the most intriguing of our newly detected streams.

\section{Conclusions}
\label{sec8}

We have used SDSS/SEGUE data from the seventh SDSS public 
data release to search for halo streams
in a 2 kpc sphere centered on the Sun. Using the cataloged values for $\log$ g,
$g-i$ color and [Fe/H], together with distance estimates based on the photometric
parallax relation from \citet{ive08}, we assembled a sample of 22,321 subdwarfs
with [Fe/H] $\leq-0.5$, excluding main-sequence turnoff stars. A comparison with
fiducial sequences for 12 globular clusters from An et al. (2008) suggests that
our distances are accurate to within systematic errors of $<5\%$, although they are
much less constrained on the high-metallicity end of the photometric parallax relation.

We divided our sample into four sub-samples, equally spaced by $0.5$ dex in
metallicity. Assuming a spherical potential, we searched for stellar streams in
each sub-sample in a space spanned by the quantities
$V_\text{az}=\sqrt{V^2+W^2}$, $V_{\Delta
\text{E}}=\sqrt{U^2+2(V_\text{az}-V_\text{LSR})^2}$, and
$\nu=\arctan\frac{V+V_\text{LSR}}{W}$. These quantities are approximations for a
star's azimuthal velocity or angular momentum, eccentricity, and orbital
angle with respect to the positive $z$-axis. 

Our basic results can be summarized as follows:
\begin{enumerate}

        \item Our sample is dominated by stars on disk-like orbits; the fraction
        of these stars with orbital inclinations between $75^\circ$ and
        $105^\circ$ is 81\% (80\%) for stars with $-1.0\leq\text{[Fe/H]}<-0.5$,
        46\% (44\%) for stars with $-1.0\leq\text{[Fe/H]}<-0.5$, and remains
        constant at 23\% (16\%-17\%) for all stars more metal-poor than
        [Fe/H]$=-1.5$ (the number in parenthesis gives the fraction of stars on
        prograde orbits). This implies that beyond [Fe/H]$\simeq-1.5$ the
        fraction of thick-disk stars remains constant. 

        \item In the metallicity range $-1.0\leq\text{[Fe/H]}<-0.5$ it is
        difficult to identify substructure among the thick-disk stars, because the smooth
        component dominates. As the fraction of thick-disk stars decreases, we detect
        signals of the stream first described by \citet{ari06} and \citet{helm06}. 

        \item We find an overdensity of stars
        moving with disk-like kinematics, but too metal-poor to belong to the
        classical metal-weak thick disk, which should not extend beyond
        [Fe/H]$\simeq-2.2$ \citep{chi00}. We interpret this clump, named `C4',
        as a tidal stream accreted on an orbit in the plane of the protodisk.
        Halo streams on such orbits are predicted both from numerical
        simulations of spiral galaxy formation \citep{quinn86,quinn93,abadi03b}
        and from the considerable (20\%) fraction of halo stars on low-eccentricity
        orbits \citep{chi00}. 

        \item We confirm the existence of previously
        detected halo streams: The `RAVE' stream, which was discovered in data
        from the first RAVE data release \citepalias{kle08}, the stream `S$_3$'
        found by \citet{det07}, which crosses the solar neighborhood from the
        direction of the South Galactic Pole on a diagonally retrograde orbit,
        and the `H99'-stream discovered by \citet{helm99} at high angular
        momentum. The latter is the most significant stream in our sample
        ($\sigma\approx12.0$), which explains why \citet{helm99} found it
        in a sample of only 275 stars. However, even with our much larger sample
        size the number of stars that belong to this stream is only
        approximately doubled. This speaks against the conclusion of
        \citet{helm99}, that as much as 10\% of the nearby halo stars originated
        from a single progenitor.
	
        The `S$_3$' stream lies very close to two other features, which we labeled `C2' and `C3', in 
        ($V_\text{az}$, $V_{\Delta\text{E}}$, $\nu$). Both posess
        similar kinematics as `S$_3$', and the small differences could well be
        explained by the effects of statistical and systematic distance errors and by the way
        we compute statistical significances, which may lead to adjacent peaks as a stellar stream enhances the
        number of stars in a certain region of phase-space. The [Fe/H] distributions of `S$_3$', `C2' and `C3' give
        strong support for the hypotheses that all three features are one single stream. The composite `S$_3$/C2/C3'
        stream shows intriguing similaraties in its spatial extent to an overdensity found by \citet{jur08} at $(R,z)\approx(9.5,0.8)$ kpc.
        If this connection is real, this would be the first time that a 6D coherent phase-space overdensity has been identified by its kinematics alone. 
        Further, because the $W$ velocities are too high for thick disk stars, we could rule out a ringlike feature in the thick disk as its origin.

        The fact that both the `RAVE' and the `S$_3$/C2/C3' stream have been found in
        two independent samples makes it very unlikely that they are ``false
        positives''.  The latter is significant at a level of
        $\sigma\gtrsim2.8$ in our data (the `S$_3$' peak, Figure~\ref{fig:sig-s3}), and in the
        data from \citet{det07}, corresponding to a confidence level of 99.5\%.
        In other words, the probability that this stream is created by chance in
        both samples is $(0.005)^2=0.0025\%$. The significance levels of the
        `RAVE' stream in the RAVE data and in the SDSS/SEGUE data are
        $\sigma\gtrsim3$ (\citetalias[Fig.~10]{kle08}) and $\sigma\approx3.0$
        (Figure~\ref{fig:sig-s3}), respectively, which also makes it very
        unlikely that both streams are created through
        Poisson noise.
	
        \item Besides the already known features, we find evidence for a large
        amount of substructure, especially in the most metal-poor bins. In
        particular, we identify one candidate for a genuine halo stream that has
        not yet been described in the literature. This stream, `C1',
        moves on highly inclined orbits nearly in the direction towards
        the North Galactic Pole. 

        \item We can roughly estimate the fraction of halo stars contained
        within the streams we detected. To obtain an upper limit, we simply
        treat all stars as halo stars that lie outside the range
        $75^\circ<\nu<90^\circ$ and outside of $V_\text{az}>0$, [Fe/H] $> -1.0$.
        This results in 4388 stars, of which 53 (1.2\%) account for the most
        well-populated peak in our sample, `C2'. The `H99' stream, with 21
        members, only represents a 0.48\% maximum fraction of halo stars. These results
        are fully consistent with statistical arguments made by \citet{gou03}
        considering the fraction of halo stars in a single stream. He puts an upper
        limit on the granularity of the halo, concluding that if the halo would be fully
        composed of kinematically cold stellar streams, then at 95\% confidence no
        single stream could contain more than a 5\% fraction of the halo stars.

        \item Metallicities are very helpful when it comes to deciding about the
        origin of a moving group. Our stream candidiates exhibit a [Fe/H]
        distribution with a single peak, indicating that their progenitor had a
        well-defined star-forming epoch. 

\end{enumerate}

This study shows the power of current and future large-scale surveys to probe
substructure in the solar neighborhood, and the Milky Way in general. 
Extensions of this technique to include additional chemical information, e.g.,
[$\alpha$/Fe] and [C/Fe] ratios, are being pursued at present.

\begin{acknowledgments}
   \textit{We thank the anonymous referee for his very thorough report which highly improved 
   our manuscript and contibuted to some of our major findings.}
    
    \textit{Funding for the SDSS and SDSS-II has been provided by the Alfred P.
    Sloan Foundation, the Participating Institutions, the National Science
    Foundation, the U.S. Department of Energy, the National Aeronautics and
    Space Administration, the Japanese Monbukagakusho, the Max Planck Society,
    and the Higher Education Funding Council for England. The SDSS Web Site is}
    http://www.sdss.org/.

    \textit{The SDSS is managed by the Astrophysical Research Consortium for the
    Participating Institutions. The Participating Institutions are the American
    Museum of Natural History, Astrophysical Institute Potsdam, University of
    Basel, University of Cambridge, Case Western Reserve University, University
    of Chicago, Drexel University, Fermilab, the Institute for Advanced Study,
    the Japan Participation Group, Johns Hopkins University, the Joint Institute
    for Nuclear Astrophysics, the Kavli Institute for Particle Astrophysics and
    Cosmology, the Korean Scientist Group, the Chinese Academy of Sciences
    (LAMOST), Los Alamos National Laboratory, the Max-Planck-Institute for
    Astronomy (MPIA), the Max-Planck-Institute for Astrophysics (MPA), New
    Mexico State University, Ohio State University, University of Pittsburgh,
    University of Portsmouth, Princeton University, the United States Naval
    Observatory, and the University of Washington.}
    
\textit{T.C.B. and Y.S.L. acknowledge partial support
from grants PHY 02-16783 and PHY 08-22648; Physics Frontier
Center/Joint Institute for Nuclear Astrophysics (JINA).
P.R.F. acknowledges support through the Marie Curie Research Training Network
ELSA (European Leadership in Space Astrometry) under contract
MRTN-CT-2006-033481.}

\textit{P.R.~F. acknowledges support through the Marie Curie Research Training Network ELSA (European Leadership in Space Astrometry) under contract MRTN-CT-2006-033481.}
\end{acknowledgments}

\appendix
\section{Properties of Stream Members}

Table~\ref{tab:APs} lists the identifications and stellar atmosperic parameters from the SSPP for stars that we have assigned to individual
streams.  The errors on [Fe/H], Teff, and log g are internal estimates, obtained
from averaging of multiple techniques.  In the cases of independently observed
stars, results have been averaged. Note that in the current version of the SSPP the astrophysical parameters have been updated to slightly different values with smaller intrinsic errors than the ones used in this study which are listed below.

Table~\ref{tab:Ph} lists the SDSS photometry, distance estimates and errors, heliocentric radial
velocities, and derived $U, V, W$ used in this study.

\begin{deluxetable}{rrrrrrrrrrr}
\tablecaption{Putative Members of the Identified Streams: Astrophysical Parameters\label{tab:APs}}
\tabletypesize{\tiny}

\tablehead{\colhead{STREAM} & \colhead{IAU NAME} & \colhead{PLATE-MJD-FIBER} & \colhead{$\alpha$} & \colhead{$\delta$} & \colhead{[Fe/H]} & \colhead{$\sigma_\text{[Fe/H]}$} & \colhead{T$_{\rm eff}$} & \colhead{$\sigma_{\text{T}_{\rm eff}}$} & \colhead{log g} & \colhead{$\sigma_\text{log g}$} \\ 
\colhead{} & \colhead{} & \colhead{} & \colhead{($^\circ$)} & \colhead{($^\circ$)} & \colhead{} & \colhead{} & \colhead{(K)} & \colhead{(K)} & \colhead{} & \colhead{} } 

\startdata
C1 & SDSS J003929.56+143550.1 & 1896-53242-317 & 9.8731813 & 14.5972557 & -1.532 & 0.032 & 5626 & 37 & 4.155 & 0.156 \\
C1 & SDSS J004044.06+143713.5 & 1896-53242-275 & 10.1835690 & 14.6204119 & -1.821 & 0.008 & 4618 & 89 & 4.572 & 0.294 \\
C1 & SDSS J004209.94+004735.8 & 1905-53613-379 & 10.5414352 & 0.7932640 & -1.643 & 0.008 & 4886 & 136 & 3.570 & 0.330 \\
C1 & SDSS J004209.94+004735.8 & 1905-53706-372 & 10.5414352 & 0.7932640 & -1.888 & 0.078 & 4749 & 43 & 3.935 & 0.363 \\
C1 & SDSS J011202.72+244140.1 & 2040-53384-296 & 18.0113449 & 24.6944771 & -1.457 & 0.054 & 5585 & 32 & 3.965 & 0.180 \\
C1 & SDSS J012737.07+385706.2 & 2042-53378-555 & 21.9044685 & 38.9517097 & -1.209 & 0.082 & 5479 & 75 & 4.489 & 0.069 \\
C1 & SDSS J012737.07+385706.2 & 2043-53351-114 & 21.9044685 & 38.9517097 & -1.345 & 0.071 & 5471 & 69 & 4.501 & 0.083 \\
C1 & SDSS J012823.46+375742.5 & 2062-53381-002 & 22.0977440 & 37.9618034 & -1.417 & 0.061 & 5121 & 50 & 4.431 & 0.151 \\
\enddata

\tablecomments{The last two entries in the bottom of the list are stars that
appear in both `S$_3$' and `C2', and both `S$_3$' and `C3', respectively. The
reason is that our selection criteria for `C2', `C3' and `S$_3$' show some
overlap. Also, we already noted in \S\ref{sec:S3C2} and \S~\ref{sec:C3C4}
that a connection between these streams probably exists.}

\end{deluxetable}

\begin{deluxetable}{rrrrrrrrrrrrr}

\tabletypesize{\tiny}
\tablecaption{Putative Members of the Identified Streams: Photometry and Velocities\label{tab:Ph}}

\tablehead{\colhead{STREAM} & \colhead{IAU  NAME} & \colhead{$u$} & \colhead{$g$} & \colhead{$r$} & \colhead{$i$} & \colhead{$z$} & \colhead{$d$} & \colhead{$\sigma_d$} & \colhead{$v_\text{rad}$} & \colhead{$U$} & \colhead{$V$} & \colhead{$W$} \\ 
\colhead{} & \colhead{} & \colhead{(mag)} & \colhead{(mag)} & \colhead{(mag)} & \colhead{(mag)} & \colhead{(mag)} & \colhead{(kpc)} & \colhead{(kpc)} & \colhead{(km~s$^{-1}$)} & \colhead{(km~s$^{-1}$)} & \colhead{(km~s$^{-1}$)} & \colhead{(km~s$^{-1}$)} } 

\startdata
C1 & SDSS J003929.56+143550.1 & 16.845 & 15.809 & 15.329 & 15.132 & 15.116 & 0.708 & 0.044 & -110.1 & 50.3 & -25.0 & -57.6 \\
C1 & SDSS J004044.06+143713.5 & 19.647 & 17.808 & 16.987 & 16.655 & 16.459 & 0.798 & 0.032 & -112.5 & 52.0 & -30.4 & -58.7 \\
C1 & SDSS J011202.72+244140.1 & 18.173 & 17.193 & 16.721 & 16.543 & 16.501 & 1.421 & 0.080 & -152.8 & 115.7 & -4.9 & -57.8 \\
C1 & SDSS J012737.07+385706.2 & 17.568 & 16.406 & 15.943 & 15.733 & 15.684 & 0.966 & 0.056 & -58.15 & -106.6 & -1.8 & -59.9 \\
C1 & SDSS J012823.46+375742.5 & 20.032 & 18.645 & 18.039 & 17.814 & 17.738 & 1.933 & 0.114 & -95.7 & -52.6 & -14.5 & -62.5 \\
C1 & SDSS J013249.36-091356.1 & 18.394 & 17.225 & 16.781 & 16.602 & 16.536 & 1.712 & 0.170 & 59.9 & -102.8 & -16.3 & -57.5 \\
C1 & SDSS J014657.63-002829.4 & 20.847 & 19.088 & 18.343 & 18.044 & 17.879 & 1.854 & 0.107 & 49.2 & -110.2 & -2.3 & -47.8 \\
C1 & SDSS J021054.90+224058.0 & 19.032 & 17.664 & 17.037 & 16.805 & 16.738 & 1.204 & 0.080 & 37.4 & -144.5 & -13.9 & -56.8 \\
\enddata

\tablecomments{The last two entries in the bottom of the list are stars that
appear in both `S$_3$' and `C2', and both `S$_3$' and `C3', respectively. The
reason is that our selection criteria for `C2', `C3' and `S$_3$' show some
overlap. Also, we already noted in \S\ref{sec:S3C2} and \S~\ref{sec:C3C4}
that a connection between these streams probably exists.}

\end{deluxetable}

\clearpage
\end{document}